\catcode`\@=11
\def\gsim{\ifmmode{\mathrel{\mathpalette\@versim>}}
    \else{$\mathrel{\mathpalette\@versim>}$}\fi}
\def\lsim{\ifmmode{\mathrel{\mathpalette\@versim<}}
    \else{$\mathrel{\mathpalette\@versim<}$}\fi}
\def\@versim#1#2{\lower 2.9truept \vbox{\baselineskip 0pt \lineskip
    0.5truept \ialign{$\m@th#1\hfil##\hfil$\crcr#2\crcr\sim\crcr}}}
\catcode`\@=12
\def\lsun{\hbox{L$_\odot$}}

\def\msun{\hbox{M$_\odot$}}
\def\ha{\hbox{H$\alpha$}}
\def\hb{\hbox{H$\beta$}}

\def\hii{\hbox{H~{\scriptsize II}}}
\def\cii{\hbox{C~{\scriptsize II}]}}
\def\ni{\hbox{[N~{\scriptsize I}]}}
\def\nii{\hbox{[N~{\scriptsize II}]}}
\def\sii{\hbox{[S~{\scriptsize II}]}}
\def\oi{\hbox{[O~{\scriptsize I}]}}
\def\oii{\hbox{[O~{\scriptsize II}]}}
\def\oiii{\hbox{[O~{\scriptsize III}]}}

\documentstyle[11pt,aaspp4]{article} 

\begin{document}

\newcommand{\esca}{erg s$^{-1}$ cm$^{-2}$ \AA$^{-1}$}

\title{THE MINI AGN AT THE CENTER OF THE ELLIPTICAL GALAXY NGC~4552 WITH
    HST\footnotemark}

\footnotetext{Based on observations with the NASA/ESA Hubble Space
    Telescope, obtained at the Space Telescope Science Institute, which
    is operated by AURA, Inc., under NASA Contract NAS 5-26555.}

\author{Michele Cappellari\altaffilmark{2,3}, Alvio
    Renzini\altaffilmark{3}, Laura Greggio\altaffilmark{4,5},
    Sperello di Serego Alighieri\altaffilmark{6,7}, Lucio M.
    Buson\altaffilmark{8}, David Burstein\altaffilmark{9} \&
    Francesco Bertola\altaffilmark{2}}

\altaffiltext{2}{Dipartimento di Astronomia,
    Universit\`a di Padova, Padova, Italy}
\altaffiltext{3}{European Southern Observatory,
    Garching bei M\"unchen, Germany}
\altaffiltext{4}{Dipartimento di Astronomia,
   Universit\`a di Bologna, Bologna, Italy}
\altaffiltext{5}{Sternwarte der Universit\"at
    M\"unchen, M\"unchen, Germany}
\altaffiltext{6}{Osservatorio di Arcetri, Firenze, Italy}
\altaffiltext{7}{Centro Galileo Galilei, Islas Canarias, Espana}
\altaffiltext{8}{Osservatorio di Capodimonte, Napoli, Italy}
\altaffiltext{9}{Department of Physics \& Astronomy,
    Arizona State University, Tempe, AZ, USA}

\begin{abstract}

The complex phenomenology shown by the UV-bright, variable spike first
detected with the Hubble Space Telescope (HST) at the center of the
otherwise normal galaxy NGC~4552 is further investigated with both HST
imaging (FOC) and spectroscopy (FOS). HST/FOC images taken in 1991,
1993, and 1996 in the near UV have been analyzed in a homogeneous
fashion, showing that the central spike has brightened by a factor $\sim
4.5$ between 1991 and 1993, and has decreased its luminosity by a factor
$\sim 2.0$ between 1993 and 1996. FOS spectroscopy extending from the
near UV to the red side of the optical spectrum reveals a strong UV
continuum over the spectrum of the underlying galaxy, along with several
emission lines in both the UV and the optical ranges. In spite of the
low luminosity of the UV continuum of the spike
($\sim3\times10^5\lsun$), the spike is definitely placed among AGNs by
current diagnostics based on the emission line intensity ratios, being
just on the borderline between Seyferts and LINERs. Line profiles are
very broad, and both permitted {\it and} forbidden lines are best
modelled with a combination of broad and narrow components, with FWHM of
$\sim 3000$ km s$^{-1}$ and $\sim 700$ km s$^{-1}$, respectively. This
evidence argues for the variable central spike being produced by a
modest accretion event onto a central massive black hole (BH), with the
accreted material having possibly being stripped from a a star in a
close fly by with the BH. The 1996 broad H$\alpha$ luminosity of this
mini-AGN is $\sim 5.6\times 10^{37}$ erg s$^{-1}$, about a factor of two
less than that of the nucleus  of NGC~4395, heretofore considered to be
the faintest known AGN. Combining all observational constraints, we
estimate the mass of the BH at the center of NGC~4552 to be in the range
between $3\times 10^8$ and $2\times 10^9 \lsun$. The relevance for the
demography of BHs in galaxies of the high (HST) resolution imaging and
spectroscopy capable of revealing an extremely low level AGN activity in
normal galaxies is briefly discussed.

\end{abstract}



\section{INTRODUCTION}

With the intent of studying the stellar populations of early-type
galaxies, in 1993 we obtained with the Faint Object Camera (FOC) on the
Hubble Space Telescope (HST) images in several ultraviolet bands of the
central regions of the elliptical galaxies NGC~1399 and NGC~4552 and of
the bulge of the Sb galaxy NGC~2681. A point-like source -- that we term
{\it spike} for short -- was evident at the center of both NGC~4552 and
NGC~2681, with their photometric profile being indistinguishable from
the PSF of the aberrated, pre-COSTAR HST (Bertola et al.\ 1995).

By comparison with another FOC image of NGC~4552 taken in 1991 (Crane et
al.\ 1993), we soon realized that this spike had increased its
luminosity in the $U$ band (F342W) by a factor $\sim 7\pm1.5$ between
the two epochs, reaching $\sim 10^6\lsun$ (Renzini et al.\ 1995,
hereafter Paper I). A second point-like source is also present in the
1991 image, $\sim 0\farcs14$ from the central spike and with nearly the
same luminosity. While both sources were detected at the $\sim 4\sigma$
level in the 1991 image, the offcenter source was not detectable in the
1993 image at the $\sim 2.5\sigma$ level.

In Paper I we discussed several possible interpretations for the
occurrence of such an ultraviolet {\it flare} at the center of NGC~4552,
favoring an accretion event onto a central massive black hole (BH). The
accreted material could have been tidally stripped from a star in a
close fly-by with the BH, though alternative ways of feeding the BH were
also considered. If the flaring spike was due to accretion onto a BH,
its spectrum should  show prominent, rotationally broadened emission
lines, typical of an accretion disk (Renzini 1995). To check this
expectation we then obtained further HST observations, both for
spectroscopy with the Faint Object Spectrograph (FOS), and for imaging
with FOC to follow the subsequent photometric evolution of the two
sources. Indeed, with the superior performance of the post-COSTAR
HST/FOC one could check whether any sign remained of the offcenter spike
that was present only in the 1991 F342W image, and for which we had no
obvious interpretation to offer (Paper I; Renzini 1995). The HST
observations with both FOC and FOS were done on May 23, 1996, and the
results are reported in this paper. These observations confirm the
variability of the central source, the disappearance of the offcenter
one, and lend conclusive support to the BH accretion interpretation of
the event: indeed, very broad emission lines (FWHM $\simeq 3000$ km
s$^{-1}$, for both permitted {\it and} forbidden lines!) characterize
the spectrum of the variable nucleus of NGC~4552, and we argue that this
object is likely to be the least luminous AGN so far discovered.

In terms of its physical properties, NGC~4552 (M89) is a rather typical
giant elliptical galaxy in the Virgo cluster, having an absolute B
magnitude of $M_{\rm B}$=$-$20.2 (adopting a distance of 15.3 Mpc, cf.
Faber et al.\ 1997), Its physical properties fit well the fundamental
plane  (Bender, Burstein \& Faber 1992) as well as the Mg$_2-\sigma$
relation (Mg$_2$=0.343 mag, $\sigma = 275$ km s$^{-1}$; Davies et al.\
1987; Bender, Burstein \& Faber 1993).  Even its far-UV properties are
typical of gE galaxies of its Mg$_2$ strength (Burstein et al.\ 1988;
Brown et al.\ 1995, 1997).

H$\alpha$ emission extends from the center of NGC~4552 out to
$\sim 2.5$~kpc (Trinchieri \& di Serego Alighieri 1991; Macchetto et al.\
1996), likely the result of a recent accretion of a small gas rich
satellite, a rather common finding among giant ellipticals.  Possibly
related to this event are the inner dust patches first recognized by van
Dokkum  \& Franx (1995) on early WFPC images, while the  presence of a
circumnuclear, dusty  ring-like structure (0\farcs2 in diameter)
encircling the central bright spike has been reported by Carollo et al.\
(1997a; cf. Section~\ref{sec:colormap} of the present paper).  This
structure is virtually identical to a feature seen at the center of
NGC~4261 (Jaffe et al.\ 1996).  Yet, ground based spectroscopy with
2\arcsec$\times$4\arcsec\ apertures of the center of NGC~4552 reveals a
spectrum dominated by broad absorption lines typical of the old stellar
populations of ellipticals, and no obvious signs of nuclear activity
(Davies et al.\ 1987; Ho, Filippenko, \& Sargent 1995).

In X-ray properties, NGC~4552 has an unresolved ``hard'' X-ray source
and a more diffuse, ``soft'' (i.e. thermal) X-ray source of comparable
luminosity (L$_{X}\sim$10$^{40.5}$ erg s$^{-1}$), as reported by
Schlegel et al.\ (1998). This is well in the range of X-ray luminosities
of ellipticals of similar optical luminosity (see e.g. Pellegrini 1994),
with the emission being probably dominated by the hot interstellar
medium. NGC~4552 also hosts a weak, compact radiosource that is variable
on a time scale of several years (Jenkins 1982; Wrobel 1991). Therefore,
several hints of BH-related nuclear activity have existed before our
detection of the time-variable spike.  Yet, NGC~1399 is a much stronger
radiosource (Bolton, Wright, \& Savage 1979), but our 1993 FOC images
reveal no central spike.

As argued  in Paper I, the UV high resolution imaging of the central
regions of galactic spheroids -- that currently can only be done with
HST -- can offer an attractive way of recognizing the presence of a
massive BH. The traditional way of revealing the presence of central BHs
in otherwise `normal' galaxies is to resort to the kinematical behavior
of stars and gas in their central regions, as very high velocities of
{\em bound} matter within a tiny volume of the galaxy is the
unmistakable signature of a massive compact object. As a consequence,
evidence for central BHs in normal or weakly active galaxies has been
extensively sought by means of stellar dynamics diagnostics (Kormendy \&
Richstone 1995; Bender et al.\ 1996; van der Marel et al.\ 1997;
Kormendy et al.\ 1997; and references therein). The existence of a
massive dark object (MDO, an euphemism for BH) at the center of {\em
virtually all} elliptical and spiral bulges with a mass $M_\bullet\simeq
(0.002-0.006)\times M_{\rm spheroid}$ is proposed by Kormendy \&
Richstone (1995) and Faber et al.\ (1997) (see Ho 1998 for a recent
review). Dynamical modelling of a sample of 36 nearby galaxies appear to
require the presence of a central MDO in at least 30 of them, with
$M_\bullet\simeq 0.005\times M_{\rm spheroid}$ (Magorrian et al.\ 1998).
On this basis, we would expect the center of NGC~4552 to house a BH of
mass $10^8$ to $10^9$ \msun.

Gas dynamics diagnostics has also been used, including studies of the
optical emission lines originating in material orbiting the putative BH,
with classical examples being the nuclear disk in in M87 (Ford et al.\
1994), the spectacular H$_2$O megamaser at the center of NGC~4258
(Miyoshi et al.\ 1995), and the time-varying profile of the H$\alpha$
emission at the center of the LINER galaxy NGC~1097 (Eracleous et al.\
1995).

Besides the 1991 and 1993 FOC data and our May 1996 HST FOC and FOS data
in this paper we also use data retrieved from the STScI archive, namely
the January 1997 FOS spectra and WFPC2 images of Faber et al.\ (1997).
The photometric evolution of the spike as seen in FOC images is
presented in Section 2.  In Section 3 the FOS spectroscopic observations
are analyzed in detail. In Section 4 we discuss several implications of
our findings. In particular, we compare the active nucleus in NGC~4552
to the previously known faintest AGN, i.e., NGC~4395 (Filippenko, Ho, \&
Sargent 1993), we speculate as to why the active nucleus is so faint,
and why the 1991 offcenter spike has virtually disappeared. We then try
to set constraints on the mass of the central BH, we compare the central
spike in NGC~4552 to those seen in other nearby galaxies, and finally we
discuss how the NGC~4552 phenomenon can be used to diagnose the presence
of a central massive BH in other galaxies.  Our conclusions are
summarized in Section 5, while in the Appendix we give the technical
details of our analysis of HST data.

\section{FOC IMAGING}

\subsection{Observations}

Pre-COSTAR FOC observations of NGC~4552 include a F342W and a F502M
frame obtained on July 19, 1991 (Crane et al.\ 1993), and our own
subsequent images obtained on November 28, 1993 in four adjacent UV
passbands (F175W, F220W, F275W, F342W). The 1991 FOC/96 image was
recorded in the 512$\times$512 unzoomed mode yielding a field of view
(FoV) of 11\arcsec$\times$11\arcsec, whereas the zoomed 512$\times$1024
mode (with a 22\arcsec$\times$22\arcsec\ FoV) was adopted for the 1993
observations. We then observed NGC~4552 for a third time on May 23, 1996
in the 512$\times$512 unzoomed mode (7\arcsec$\times$7\arcsec\
COSTAR-Corrected FoV) with the three UV filters (F175W, F275W and
F342W). A summary of the instrumental configurations and exposure times
is given  in Table~\ref{tab:obs_log}.   The FOC observations have been
calibrated using the standard STScI pipeline software. The most recent
calibration files have been used when dearchiving and recalibrating the
older data from the HST Archive.

\subsection{Position of the Flare}

The location of the central and offcenter spikes with respect to the
photometric center of the galaxy is determined using the ELLIPSE task in
IRAF/STSDAS over the F342W images obtained in 1991, 1993 and 1996. The
result is shown in Figure~\ref{Identificazione_Spike} with superimposed
a contour plot of the 1991, 1993 and 1996 images (3 pixel boxcar
smoothed) of the very central region. Here the ``X'' denotes the
position of the center of the 1\arcsec\ semimajor axis elliptical
annulus. Note that the contour levels do not correspond to the same
surface brightness on the different images, since they are only intended
to display the position and structure of the spikes.  The elliptical
shape of the spike in the 1993 image is due to the zoomed mode of the
original FOC image (see Table~\ref{tab:obs_log}), while the smaller size
of the spike in the 1996 image is an effect of the deployment of the
COSTAR corrective optics.

The three plots have the same scale and orientation: north is up and
east at the left. It can be seen that in 1991 the center of the galaxy
coincides with one of the two point-sources seen in that image (cf.
Paper I), while in the 1993 and 1996 images it coincides within
$\sim0\farcs02$ with the sole spike seen. This coincidence between the
center of the galaxy and the spike has also been verified on the images
obtained with the other UV filters, although with lower accuracy, due to
the lower S/N.  We note that while the formal errors given by ELLIPSE on
the galaxy center position are negligible, other factors, such as the
accuracy of FOC flat-fielding and geometric corrections play a role in
the determination of the center, so the small offset of the central
spike with respect to the galaxy photometric center is not significant.
On the other hand, the relative offset of the two point sources in the
1991 image is significant and reliably determined.

\subsection{Photometric Modeling\label{sec:phot_model}}

To measure the luminosity of the central spikes one has to model the
underlying surface brightness distribution of the galaxy. For this
purpose we adopt the convenient parametrization now known as the
``nuker'' law (Lauer et al.\ 1995):
\begin{equation}
    \label{eq:nuker}
    I(r) = I_{\rm b}\; 2^{\frac{\beta-\gamma}{\alpha}}
        \left(\frac{r}{r_{\rm b}}\right)^{-\gamma}
        \left[1 + \left(\frac{r}{r_{\rm b}}\right)^{\alpha}\right]
        ^{-\frac{\beta-\gamma}{\alpha}},
\end{equation}
where $\gamma$ measures the steepness of the inner profile ($I(r)
\propto r^{-\gamma}$ for $r \ll r_{\rm b}$), $\beta$ is the steepness of
the outer profile  ($I(r) \propto r^{-\beta}$ for $r \gg r_{\rm b}$),
$\alpha$ is related to the sharpness of the transition and $I_{\rm b}$
is a scale factor. To this profile we will add the contribution of an
unresolved central point source ($c_s$), that represents the predicted
flux of the {\em spike}. The derived parameters
for each of the FOC epochs are given in Table~2.
A central spike feature is present in a fair
number ($\sim$25\%) of the early-type galaxies modeled with a nuker law
(Lauer et al.\ 1995, Carollo et al.\ 1997a,b). However, only for NGC~4552
 multiple epoch UV images and a high resolution
spectrum in the UV  are available, so further data are required to
asses  whether there is any
similarity between the spike in NGC~4552  and these other spikes.

The comparison of the FOC images obtained at the three epochs is
complicated by two factors.  First, each FOC image was obtained under
different conditions:  The 1991 image is pre-COSTAR, FOC-standard; the
1993 image is pre-COSTAR, FOC-zoomed, while the 1996 image is
post-COSTAR, FOC-standard. Second, the 1993 images exceed the linearity
limit of the FOC in some of the filters. As such, we have re-calibrated
all the images in a self-consistent manner, including all required
correction factors for PSF and sensitivity differences (zoom/non-zoomed
modes and COSTAR) and nonlinearity effects.

In so doing, we are aided by the fact that the intrinsic luminosity
profile of the galaxy in a given FOC filter has to be the same for all
epochs.  In contrast, we permit the luminosity of the spike to vary.
Owing to the smaller field size of the post-COSTAR 1996 image, only the
shape of the inner galaxy luminosity profile (i.e., the $\alpha$,
$\gamma$ and $r_{\rm b}$ parameters) can be measured from the 1996
non-aberrated (linear) images.  The slope of the outer luminosity
profile ($\beta$) is better be measured from the 1993 images, which
cover the largest field of view and from which the dark rate can be
reliably determined within the shade of the coronographic fingers.  In
this way, for each FOC filter a unique two-dimensional luminosity
distribution for NGC~4552 is obtained (as described by the parameters
$\alpha$, $\beta$, $\gamma$ and $r_{\rm b}$.  A two-dimensional model is
required in order to apply the PSF-related corrections to the data.)  We
confirm that, when convolved in the appropriate manner for each FOC
instrumental configuration, the galaxy luminosity profile accurately
reproduces the observed profiles.  What can, and does differ for each
passband and epoch is the scale factor $I_{\rm b}$ and the excess counts
$c_{\rm s}$ from the central spike (plus the second, off-center spike in
the 1991 image).  In practice, an iterative procedure was used to derive
the galaxy and spike parameters, in the following manner:

\begin{enumerate}

\item A dark rate of $7\times10^{-4}$ counts s$^{-1}$ (Nota et al.\
1996) is initially assumed and subtracted from the 1996 image. Since
this is not the true background, this assumption will affect the
determination of the outer Nuker law parameter $\beta$.

\item The initially background--subtracted image is then iteratively
proceed through the following four steps (a---d) to provide a best-fit
estimate of the Nuker-law parameters $\alpha$, $\gamma$ and $r_{\rm b}$.
A Nelder-Mead Simplex algorithm is used for the first iterations, when
the parameters are still far from the exact solution, while a
Levenberg-Marquardt algorithm is used when the solution is approached to
accelerate convergence and to estimate the errors (see Press et al.\
1992). Since the ellipticity measured from our images is very small
($\varepsilon\lsim0.03$) and the position angle can not be reliably
determined, $\varepsilon\equiv0$ was set for all models.

\begin{enumerate}

\item A two dimensional synthetic model is made of the galaxy surface
brightness distribution with a photometric profile described by
Equation~(\ref{eq:nuker}).

\item A certain number of counts $c_s$ is added to the central pixel of
this model to simulate the point-like central spike.

\item This synthetic image is then convolved with the appropriate PSF
taken from the STScI library, and the effects of  nonlinearity are
included (described in Appendix~\ref{sec:nonlin}).

\item The photometric profile of the {\em simulated} image of the galaxy
is fit, on a $\log-\log$ scale, to the photometric profile of the {\em
observed} image by minimizing $\chi^2$, and an improved full set of
Nuker law and spike parameters is determined.

\end{enumerate}

\item After the ``inner'' Nuker law parameters $\alpha$, $\gamma$ and
$r_{\rm b}$ are determined from the previous step, the ``outer''
parameter $\beta$, the scale factor $I_{\rm b}$ and the counts from the
spike $c_s$  are iteratively fit to the 1993 image, following the
procedure defined in Steps 2a--2d.

\item  The value of $\beta$ determined from the 1993 image is then used
to iteratively constrain the dark count rate of the 1996 image, by
requiring this image to yield the same value of $\beta$. During this
step all other Nuker law parameters are kept fixed as determined by
Step~2.

\item The resulting full set of parameters ($\alpha$, $\gamma$, $r_{\rm
b}$ and $\beta$) as determined from Steps 2--4 are then used as input
and the process is begun again at Step 2 until convergence is reached
for all parameters, including the spike count-rate, for the 1993 and
1996 FOC images.

\end{enumerate}

The result of this iterative procedure is the determination of the Nuker
parameters for all images of NGC~4552 in the three filters F175W, F275W,
and F342W, plus the spike counts in the same bands for both the 1993 and
1996 images. The counts for the central spike in the F342W image of 1991
are then estimated (together with the scaling factor) as a free
parameter of a fit whose Nuker law parameters are now forced to the
values determined for both the 1993 and 1996 images.  The counts of the
off-center spike detected in our 1991 image is then separately
determined by direct PSF-fitting.

For two filters, only one observation is available --- F502W in 1991 and
F220W in 1993 --- so the above iterative procedure is not appliable. For
the F220W image we have {\em assumed} that the parameters $\alpha$,
$\gamma$ and $r_b$ of the ``Nuker'' law are the same as for the F275W
image and we only apply Step 3 of the sequence enumerated above.  In the
case of the F502M image, the contribution of the spike is negligible
relative the galaxy luminosity profile itself, so we have fit all of the
parameters using Steps 2a--2d.

\subsection{Results}

The results of the Nuker law plus central spike fits are shown in
Figure~\ref{F342W_1991_NGC4552} for all FOC images taken from 1991 to
1996, nine in all. Note that the fits have been restricted within
$r<3\arcsec$ from the center in order to better reproduce the nuclear
profile and to reduce the sensitivity to dark subtraction errors. We
have used the STScI magnitude system, defined by the relation
\begin{equation}
    m_{\rm HST}^{\rm j} = -21.1 -2.5\ \log \frac{{\rm counts}\times U_{\rm
j}}{T_{\rm j}}
\end{equation}
where $U_{\rm j}$ is the filter inverse sensitivity and $T_{\rm j}$ is
the exposure time. Determination of the inverse sensitivities is
discussed below (Sec. 2.4.1). Note however that the definition of our
photometric system is actually quite arbitrary, as the red leak of the
UV filters can be significant and hard to predict for the red spectral
energy distribution (SED) of the galaxy. However, the red-leak gives a
negligible contribution to the FOC photometry of the spike due to its very
blue spectrum.

The numerical values of the parameters determined with these models are
presented in Table~\ref{tab:phot}. The 1--$\sigma$ formal errors of the
fit parameters are computed from the covariance matrix, so they follow
from the shape of the $\chi^2$ surface near its minimum. The parameters
$I_b$ and $c_s$ in column (7) and (8) are given in units of raw counts,
as it is in this form they will be discussed later.

\subsubsection{Calibration of the Frames}

The ratio $U_{\rm j}^{-1}/U_{\rm k}^{-1}$ of the sensitivity of two
different frames obtained with the same filter is given by the following
relation
\begin{equation}
    \frac{U_{\rm j}^{-1}}{U_{\rm k}^{-1}} = \frac{I_{\rm b}^{\rm j}\
    T_{\rm k}\ {\rm scale}_{\rm k}^2}{I_{\rm b}^{\rm k}\ T_{\rm j}\
    {\rm scale}_{\rm j}^2}
\label{eq:sens_ratio}
\end{equation}
where $I_{\rm b}^{\rm k}$ is the scaling factor (Table~\ref{tab:phot}),
$T_{\rm k}$ is the corresponding exposure time
(Table~\ref{tab:obs_log}), scale$_{\rm k}=0.02250$ arcsec pixel$^{-1}$
for the pre-COSTAR observations (1991 and 1993) and scale$_{\rm
k}=0.01435$ arcsec pixel$^{-1}$ for the post-COSTAR ones (1996) (Nota et
al.\ 1996).

In Table~\ref{tab:calib} we have computed the sensitivity ratios
[Eq.~(\ref{eq:sens_ratio})] between the different observing modes by
taking the ratios of the galaxy model fluxes determined from the model
fits (Sec. 2.3). The ratio of the 1993 galaxy model to the 1991 galaxy
model depends on the sensitivity difference between the 512$\times$1024
zoomed mode (1993) relative to the 512$\times$512 standard mode (1991).
The value of $1.27\pm0.02$ we derive from this comparison is very close
to the $1.25\pm0.05$ factor tabulated in the latest FOC manual. This
comparison makes us confident that our calibration method is reliable
and accurate.

Although the 1991 and 1996 FOC frames are taken in the same observing
mode (512$\times$512 mode), we have to take into account the net
decrease in sensitivity in the 1996 frame owing to two additional COSTAR
reflections. In addition we also have to take into account any possible
sensitivity degradation over this five year period. While the
$1.45\pm0.07$ decrease (1991 to 1996) we measure is bigger than the 1.23
value we get using the SYNPHOT package within IRAF/STSDAS, this
difference is understandable in the following way:
Our method estimates the response ratios of the FOC via surface
brightness measurements on a large area of the detector. In contrast,
SYNPHOT estimates are based on aperture photometry of standard stars, which
by its very nature combines changes in the PSF and true sensitivity
variations (which both have occurred between 1991 and 1996) together
into one correction.
As a result we would expect our estimate of {\it sensitivity} decrease (1991 to
1996) to be larger than that obtained using SYNPHOT, which is what we
observe.

Our estimates of sensitivity variations among the different FOC modes
are conservative in the sense of minimizing the luminosity variation of
the spike from 1993 to 1996. If we used the SYNPHOT estimates of
sensitivity variation we would obtain a larger variation of the spike
luminosity. To convert our relative fluxes to physical units, we {\em
adopt} the STScI absolute calibration of the COSTAR corrected images (as
given by SYNPHOT) as the true value, and we rescale this sensitivity
according to the sensitivity ratios we determined above.

\subsubsection{Variation of the Spike}

Since the central spike is allowed to vary in the different models, we
can determine its variation from the data of Table~\ref{tab:phot}. The
ratio $f_{\rm j}/f_{\rm k}$ of the flux from the spike in a given band
at two different dates can be computed from the equation below:
\begin{equation}
    \frac{f_{\rm j}}{f_{\rm k}} = \frac{c_{\rm s}^{\rm j}\ I_{\rm
                           b}^{\rm k}\ {\rm scale}_{\rm j}^2}
                           {c_{\rm s}^{\rm k}\ I_{\rm b}^{\rm j}\ {\rm
                           scale}_{\rm k}^2}
\end{equation}

In Table~\ref{tab:calib} and in Figure~\ref{variazione_spike} is
presented the variation of the spike in all the bands where it has been
possible to measure it. In table~\ref{tab:fluxes} the flux from the
spike is shown in physical units. Note that, as will be discussed later,
the spectrum of the spike is rather hot. For this reason the red-leak is
negligible and we can convert from counts within one filter to \esca\ at
the effective wavelength $\lambda_j$ of the filter by using the inverse
sensitivity of that filter [$f(\lambda_{\rm j}) = {\rm counts}\ U_{\rm
j} / T_{\rm j}$].

As is evident, the center spike in NGC~4552 increased in luminosity by a
factor of 4.5 from 1991 to 1993 confirming the results presented in
Paper I. Moreover it appears that the spike faded from 1993 to 1996,
decreasing in luminosity by about a factor of 2.0 at all observed
wavelengths (1700--3500 \AA).

\section{FOS SPECTROSCOPY}

\subsection{Observations}

FOS spectra of the central spike of NGC~4552 were obtained on May 23,
1996 as part of the same HST/GO program (GO 6309), using the
0\farcs21$\times$0\farcs21 square aperture (0.25--PAIR), the G270H,
G650L and G780H gratings and the red FOS detector (FOS/RD).  From the
last FOS peak-up target acquisition stage we have verified that the
spike was correctly centered to an accuracy of less than 0\farcs04 in
the FOS aperture.  A subsequent FOS spectrum was obtained on January 16,
1997 on a different GO program (6099, PI: S.M. Faber) using the G570H
grating and the same aperture and detector of our observations. We have
retrieved this spectrum from the STScI archive to compare it with our
data. A log of these observations is presented in
Table~\ref{tab:fos_log}.

All spectra have been calibrated by the standard STScI data pipeline.
Note that the latest ``Average Inverse Sensitivity'' (AIS) method was
used in the calibration. If more than one spectrum was available a
weighted average has been used to compute the resulting average
spectrum. For each output spectrum we also computed a companion ``error
spectrum,'' by properly combining the pipeline-supplied 1-$\sigma$ error
spectra. These error spectra are needed for the modeling of the spectral
features described below.

\subsection{The 1996 UV/Optical FOS spectrum}

\subsubsection{The 1996 UV-Dominated Continuum of the Spike}

In Figure~\ref{Spettro_Completo_Spike_N4552} we compare the 1996 merged
FOS spectra of the central 0\farcs21$\times$0\farcs21 region of NGC~4552
with a composite spectrum meant to represent the spectrum of the inner
$r=7\arcsec$ of this galaxy.  The composite spectrum is obtained by
combining the IUE spectrum of NGC~4552 (Burstein et al.\ 1988) with the
low resolution (20 \AA) optical spectrum of NGC~4649 (Oke et al.\ 1981),
which is known to have a UV and optical SED very similar to that of
NGC~4552 (cf. Burstein et al.\ 1988).

Normalizing at V magnitude, we see that this composite spectrum for the
inner 7\arcsec\ of NGC~4552 is a very good match in overall spectral
energy distribution for the FOS spectrum, with two notable exceptions:
First, the FOS spectrum shows strong emission lines that are absent in
the composite spectrum.  Second, the UV SED  of the FOS spectrum is far
stronger than the IUE SED shortward of 3200 \AA. The SED of the spike
alone is obtained by subtracting the V-mag normalized 7\arcsec\ IUE
spectrum from the 0\farcs21$\times$0\farcs21 FOS spectrum
(Figure~\ref{Continuo_Spike_N4552}).  In the same figure we overplot the
fluxes measured for the spike using the nominal filter sensitivities for
the 1996 FOC observations, as well as blackbody SEDs for various
temperatures. It is evident that both the FOC fluxes and the FOS SED
agree well, and that together they indicate a temperature of
$T\sim15000$ K for the spike in 1996, if a thermal origin for the UV
flux is assumed. In turn, this implies a bolometric luminosity of
$\sim3\times10^5$ \lsun\ for the spike (at a distance of 15.3 Mpc).

\subsubsection{Defining The Emission Lines}

The May 1996 FOS spectra of the central spike in NGC~4552,
redshift-corrected, are presented in
Figure~\ref{Identificazione_Righe_N4552}. These data cover the range
2222--8500 \AA, with a gap from  3277--3540 \AA.  The most prominent
emission lines are identified. They include \cii\ $\lambda$2326, MgII
$\lambda$2800, \oii\ $\lambda$3727, \sii\ $\lambda$4072, \hb, \oiii\
$\lambda\lambda$4959, 5007, \ni\ $\lambda$5700, \oi\ $\lambda$6300,
\nii\ $\lambda\lambda$6548, 6583, \ha\ and \sii\ $\lambda\lambda$6717,
6731.

Under the reasonable assumption that the emission line spectrum is from the
spike alone, and the rest of the optical spectrum is underlying galaxy SED,
by subtracting a suitable template for the galaxy SED we can get a good
approximation to the spectrum of the spike. A suitable template spectrum
can be found from the spectral library of Ho et al.\ (1995).  After careful
search for a  ``clean'' galaxy spectrum (i.e., without apparent emission
lines), we chose the spectrum of NGC~3115 on the basis both of high S/N and
good overall SED match to NGC 4552. (Obviously we would have preferred
to use the Ho et
al.\ spectrum of NGC~4649 (as we did in
Figure~\ref{Spettro_Completo_Spike_N4552}) due to its closer similarity in
the UV spectrum and Mg2 index (Burstein et al. 1988), but unfortunately it
is of poor S/N.)  The Ho et al.\ spectrum of NGC~3115 was then subtracted
to the spectrum of NGC~4552 from the same library, and the result is shown
in Figure~\ref{template_subtraction}. The very small difference between the
two spectra testifies for the suitability of NGC~3115 as a template for the
stellar SED of NGC~4552. The small positive difference in the spectral
regions of H$\alpha$+\nii\ and \sii\ indicates that some emission is
detected in NGC~4552 even within the large 2\arcsec$\times$4\arcsec\
aperture used by Ho et al.

Finally, the spectrum of NGC~3115 has been appropriately scaled both in
flux and spectral resolution to the FOS spectra of NGC~4552, and
subtracted to it. The resulting pure emission line spectrum is then
taken as the spectrum of the spike on which we have conducted all
subsequent analyses.

\subsubsection{Diagnostics of the Spike Via Emission Line Ratios}

Integration over all recognized emission lines with reasonable S/N has
given the total flux in each of these lines, with the results being
reported in Table~\ref{tab:line_modeling}.
Figure~\ref{classificazione_spike_n4552} shows how the emission line
ratios of the narrow components for the NGC~4552 spike compare to the
distribution of Seyfert galaxies, LINERS and \hii\ regions in the
diagnostic emission line diagrams of Ho et al.\ (1997).  As is evident,
the line ratios to \ha\ of \nii, \sii\ and \oi\ definitively place the
spike in NGC~4552 among extreme AGNs, while the \oiii/\hb\ ratio falls
just on the borderline between Seyferts and LINERs. Therefore, the spike
in NGC~4552 can be either classified as a very high excitation LINER or
a very low excitation Seyfert.

\subsection{Fitting the Emission Lines}

The continuum-subtracted optical emission-line spectrum has been modeled
as a whole by applying a specifically constructed IDL procedure which
makes use of the Levenberg-Marquardt algorithm to fit a non-linear
function. As usual in the case of AGN spectra, as a first-order
approximation we have imposed on all the emission lines the following
model constraints: [i] Each permitted {\em and} forbidden line consists
of a narrow and a broad Gaussian component; [ii] the redshift of every
emission line in a given FOS setup is assumed to be the same; [iii] the
intensity ratio of the narrow and broad component is forced to be the
same for all lines; [iv] the intrinsic width of all the broad and narrow
components are constrained to be the same within a given spectrum. The
one exception to these rules is the redshift of H$\alpha$ in the 1997
G570H spectrum, which is found to be significantly different than that
of the other emission lines in this spectrum (see below).  The derived
parameters for the emission lines are given in
Table~\ref{tab:line_modeling}. The 1-$\sigma$ formal errors of the line
parameters are estimated from the covariance matrix.

We first illustrate the case of the unresolved UV multiplet
\cii~$\lambda$2326 for which we assume a linear continuum under the
emission line (i.e., no significant absorption under this feature). As
demonstrated by Figure~\ref{modeling_cii}, the \cii\ emission clearly
consists of both a narrow and a broad line component.  Given the low S/N of
the UV spectrum in this wavelength range, rather than taking a straight
difference of observed minus model spectra, to emphasize the
reality of each component we have applied a low pass filter (middle
spectrum in Figure~\ref{modeling_cii}) and a high pass filter (lower
spectrum), while fit was made on the original,
unfiltered data. This result justifies our assumption [i] above.

The fits for the optical emission lines measured in the 1996 spectra are
shown in detail in Figure~\ref{modeling_halpha_agn_like}. The region
near \hb\ and \oiii\ is shown from the G650L spectrum while the region
from \oi\ to \sii\ is shown from part of the G780H spectrum. The top
plot referring to each region/grating combination shows the spectrum as
observed, together with the spectrum of the template; the middle plot
shows the emission lines fitted to the template-subtracted spectrum; and
the bottom plot shows the residuals (observed spectrum minus model fit).
The same is shown in Figure~\ref{modelling_97} for the emission lines in
the 1997 G570H spectrum.

The models fit the data quite well, as can be seen both by eye and from
the values of the reduced chi-square $\chi^2_\nu$ given for each fit. In
each case and for both 1996 and 1997 data we see that the emission lines
require both a narrow and broad line component.

\subsection{Results from the Emission Line Measures}

An examination of the results of the emission line fits given in
Table~\ref{tab:line_modeling} permits us to draw the following
significant conclusions: [i] Satisfactory fits of the emission lines
identified in nuclear FOS spectra taken $\sim 8$ months apart can be
obtained only by resorting to a combination of broad and narrow
components for both the permitted as well as the forbidden lines.  This
result is at variance with the behavior of classical AGNs -- where the
broad component is usually present only in the permitted lines. [ii] The
emission lines are very broad, with fits indicating very high velocity
widths for both the broad (FWHM $\simeq 3000$ km s$^{-1}$) and narrow
components (FWHM $\simeq 700$ km s$^{-1}$). [iii] The shape of the
\ha+\nii\ complex has definitely changed from the May 1996 spectrum to
the January 1997 spectrum. There is indeed an obvious ``dip'' in the
middle of the H$\alpha$+\nii\ complex that is clearly not present in the
1996 spectrum. Within our procedure, a satisfactory fit can be formally
achieved only allowing for a shift to the blue of $\sim 230$~km s$^{-1}$
of the whole (narrow $+$ broad) H$\alpha$ line in the latter spectrum
compared to the former one.  This also implies a similar shift of the
H$\alpha$ components with respect to the \nii\ lines in the 1997
spectrum, while no hint of such a difference in radial velocity is seen
in the 1996 spectrum.

\section{DISCUSSION}

The phenomenology we have described in detail in the previous sections
can be summarized as follows: [i] The central spike UV luminosity has
varied by a factor of a few over timescales of order of 15 months, being
in the range from $10^5$ to $10^6$ $\lsun$. [ii] An offcenter UV spike
is present only in the 1991 data. [iii] The spectrum of the spike is
characterized by a strong UV continuum plus moderate excitation emission
lines. [iv] The emission line ratios are typical of AGNs, but unusual in
that both permitted and forbidden lines require broad and narrow
components. [v] The baricenter of the H$\alpha$ emission has changed by
$\sim 230$ km s$^{-1}$ between May 1996 and January 1997.

This complex phenomenology clearly points towards the presence of a very
weak AGN at the center of this galaxy, most likely powered by a low
level of accretion onto a central black hole (BH). Other interpretations
(a central supernova, a central starburst, ...) were already considered
unlikely, given the early evidence (Paper I, Renzini 1995). The
additional HST evidence gathered in 1996 is reasonably conclusive in
this respect.

\subsection{Is the Nucleus of NGC~4552 the Least Luminous Known AGN?}

With the adopted distance to the Virgo cluster of 15.3 Mpc and the
observed flux of the broad \ha\ emission from Table~6, the broad \ha\
luminosity is $\sim 5.6\times 10^{37}$ erg s$^{-1}$. This is a factor of
two less than the broad \ha\ luminosity of the nearest known AGN, the
Seyfert 1 nucleus of NGC~4395, and a factor of $\sim 20$ less luminous
than M81, the next faintest Seyfert 1  nucleus (Filippenko, Ho, \&
Sargent 1993). Filippenko et al.\ adopt a distance to NGC~4395 of 2.6
Mpc, and clearly claims of which of the two nuclei is the least luminous
have to cope with the uncertainty in the relative distance of the two
galaxies. It is beyond the scope of this paper to explore this aspect in
detail, but it is safe to conclude that given the current evidence
NGC~4552 is likely to harbor the intrinsically faintest known AGN, or
at least the next to the faintest AGN. Of course, allowance should also
be made for the confirmed variability of the mini-AGN in NGC~4552.

A comparison of the ultraviolet FOS spectra of the NGC~4552 spike and
the FOS spectrum of the AGN in NGC~4395 (Filippenko, Ho \& Sargent 1993)
shows that the latter contains lines of high ionization that are not
present in the former.  In particular, Filippenko et al.\ point out that
the existence of the Bowen fluorescence mechanism from OIII 3133 to HeII
3204 is a good diagnostic of the existence of high ionization in an AGN.
These two lines are notably absent in the FOS G270H spectrum of the
spike in NGC~4552, as are other high level Fe and Ne lines commonly seen
in high ionization AGN. However, the \cii\ and MgII lines have similar
intensities in the two objects, while the flux of the UV continuum is
decreasing towards shorter wavelengths in NGC~4395 in the spectrum of
Filippenko et al., but it is increasing in our spectrum of the spike in
NGC~4552.

Ground based spectroscopic observations of NGC~4395 in the \ha+\nii\
region have been obtained by Filippenko \& Sargent (1989, FS89) using a
2\arcsec\ aperture. Given the relative distances of the two galaxies,
our FOS and FS89 observations sample regions of quite similar physical
size. According to FS89, the narrow line and broad line components of
\ha+\nii\ in the nucleus of NGC~4395 have FWHM of $\sim 60$ and $\sim
800$ km s$^{-1}$, both much narrower than in the case of the mini-AGN in
NGC~4552. We conclude that the arguments developed by Filippenko et al.\
to exclude explanations other than BH accretions for the activity in
NGC~4395 are even more effective in the case of NGC~4552, which shows
more extreme AGN characteristics.

\subsection{Why is the AGN Luminosity so Low?\label{sec:low_agn}}

As already suggested in Paper I, the phenomenology of the spike in
NGC~4552 is generically consistent with the scenario in which a central,
UV-bright flare is caused by the tidal stripping of a star in a close
flyby with a central supermassive BH. From the theoretical point of
view, only the extreme case of a total disruption of a $\sim 1\,\msun$
main sequence star has been widely investigated so far (e.g., Rees 1988,
1990; Evans \& Kochanek 1989; Luminet \& Barbuy 1990; Cannizzo, Lee, \&
Goodman 1990; Kochanek 1994; Eracleous et al.\ 1995; Loeb \& Ulmer 1997;
Ulmer 1997a,b).

The frequency of such events is estimated to be of one total disruption
every $\sim 10^3-10^4$ yr in typical giant elliptical galaxies (Rees
1988, 1990).  The flare is predicted to be very bright for several years
($\sim 10^{10}\lsun$),  much brighter than the observed flare. This
indicates that if the flare in NGC~4552 was caused by a tidal stripping
in a BH-star flyby, then this flyby was rather wide, and led to only
partial stripping.  Obviously one expects wider flybys to be vastly more
frequent than the hard ones causing total disruption.  To be consistent
with the observed luminosity, only $\sim 10^{-3}\msun$ should have been
stripped, as the flare luminosity is predicted to scale with the 5/3
power of the mass of the accretion disk (Cannizzo et al.\ 1990;
Paper~I).

Tidal stripping of a star is but one possible way to feed matter to a
massive central BH at a low rate.  Other mechanisms include: a) Roche
lobe overflow from one or more stars in bound orbit(s) around the BH
(Hameury et al.\ 1994); b) accretion from a clumpy interstellar medium
near the central BH; or c) gas fed to the BH via a cooling flow within
the X-ray emitting hot interstellar medium known to exist in NGC~4552.

Some circumstantial support for accretion from a clumpy ISM may come
from NGC~4552 showing extended H$\alpha$ emission in the inner 2 kpc
(Trinchieri \& di Serego Alighieri 1991), possibly due to the recent
accretion of a dwarf galaxy. Concerning alternative c) above, we note
that the large dispersion in the X-ray luminosity of elliptical galaxies
for a given optical luminosity argues for the hot gas flow in most
ellipticals being directed outwards, i.e. being in either a supersonic
wind or subsonic outflow regime (Ciotti et al.\  1991; Renzini 1996).
However, even when gas flows out through most of the galaxy body, a
mini-inflow is likely to be established in the innermost ($r\lsim 300$
pc) regions, with inflow rates being as low as a few $10^{-3}\msun$
yr$^{-1}$ (Ciotti et al.\ 1991). Accretion onto the central BH is likely
to be intermittent (Ciotti \& Ostriker 1997), hence resulting in sizable
excursions in the associated luminosity (flickering), a well known
phenomenon among AGNs (Ulrich, Maraschi, \& Urry 1997).

It may well be that each of all of these feeding mechanisms operates
from time to time in the central regions of elliptical galaxies like
NGC~4552. Theoretical models of the various options all must end up with
assuming that an accretion disk is established, and from this point on
all options resemble each other very closely. From the observational
point of view, the available data do not allow a clear cut
discrimination among which of these alternatives is responsible for the
event in NGC~4552.

However, circumstantial evidence in favor of the tidal stripping option
comes from the noticed shift in the broad H$\alpha$ emission between May
1996 and January 1997. Tidal stripping/disruption is indeed predicted to
give rise to an {\it elliptical} accretion disk, the precession of which
results in sizable H$\alpha$ line profile variations (Eracleous et al.\
1995). This scenario (tidal stripping plus elliptical disk) has been
proposed to account for the observed variability of the broad H$\alpha$
line profile in the active nucleus of NGC~1097 (Eracleous et al.\ 1995;
Storchi-Bergmann et al.\ 1995, 1997).

\subsection{Origin of the Broad Forbidden Lines}

According to general wisdom, the onset of broad forbidden-line emission
in AGNs is prevented by the high densities existing in broad line
regions (BLR). The BLR is thought to be close to the central BH, and
most likely consisting of an accretion disk around it. The narrow-line
region (NLR)  is thought to be located further away from the central BH,
and likely consisting of lower density clouds being illuminated by the
radiation cone emerging from the disk.

We preliminarily note that a low density in the accretion disk is
demanded by the very low luminosity of the mini-AGN in NGC~4552. As
such, nothing may prevent the forbidden lines to originate from the disk
itself, rather than from a physically distinct region. If so, our
modelling of the emission lines as a unique combination of a broad and a
narrow component may be adequate as a first approximation, but may fail
to represent the actual geometry of the emitting region. If the whole
emission line radiation originates from a (thin) disk, then the gradient
in rotational velocity should be responsible for the actual line
profile, with the broadest part of it originating in the inner regions
of the disk, with a continuous transition to a narrower and narrower
emission towards its outer regions. However, given the low S/N of the
FOS spectra we do not consider worth venturing here into more
sophisticated line profile modelling based on detailed accretion disk
models.

\subsection{Setting Constraints on the Mass of the Central Black Hole}

The size of the emitting region that is responsible for the spike can be
estimated in two ways. As noticed in Section~\ref{sec:phot_model}, the
profile of the central spike in the 1996 images is consistent with that
of the PSF of the FOC. An upper limit to the size of the emitting region
has been obtained by convolving the PSF with a series of Gaussian
profiles of various widths. With Gaussian FWHM of 0\farcs03 one notices
that with the S/N corresponding to our data the profile starts to depart
appreciably from the PSF. We therefore adopt an upper limit of 0\farcs03
for the size of the emitting region, which corresponds to $R=1.1$ pc.
Assuming a rotation velocity of 1100 Km s$^{-1}$ (this is the
$\sigma_v=$FWHM$/\sqrt{8\log 2}$ measured on the G570H spectrum), a
rough upper limit to the mass of a central compact object then follows:
$M\lsim R v^2/G\simeq 2\times 10^9$ \msun.

Alternatively, the size of the broad line emitting region can be
estimated from the upper limit to the density as set by the forbidden
lines, and by the \ha\ luminosity. Following Osterbrock (1989)
\[
V=\frac{L({\rm H}\alpha)}{f\ n_{\rm e}^2\ \alpha_{{\rm H}\alpha}^{\rm
    eff}\ h\ \nu_{{\rm H}\alpha}},
\]
where $V$ is the volume occupied by the emitting gas, $L(H\alpha)$ is
the total luminosity of the broad $H\alpha$ line, $f$ is the volume
filling factor, $n_{\rm e}^2$ is the gas electron density, $\alpha_{{\rm
H}\alpha}^{\rm eff}=1.17\times10^{-13}$ cm$^3$ s$^{-1}$, $h$ is the
Planck constant and  $\nu_{{\rm H}\alpha}$ the frequency of the
$H\alpha$ line.

From the presence of the broad \sii\ forbidden lines we know that
$n_{\rm e}\lsim10^4$ cm$^{-3}$.  By assuming that the gas is uniformly
distributed within a sphere and a unity filling factor we get a lower
limit to the radius of the emitting region, i.e., $R\sim0.2$ pc. More
likely the emission comes from bound material orbiting around a central
mass, hence $M\gsim R v^2 / G\simeq 3\times 10^8$ \msun.

Finally another constraint to the size of the emitting region comes from
the variability of the spike. From the time scale $t\simeq2$ yr one can
say that the emitting region should be smaller than $R\lsim0.6$ pc,
which is of the same order of magnitude and consistent with our
previously determined values.

Our rough estimate, between $3 \times 10^8$ to $2 \times 10^9$ \msun\ is
consistent with the mass of $M\simeq 4-6\times 10^8$ \msun\ of the
supermassive central BH in NGC~4552, obtained by Magorrian et al.\
(1998), using simple {\em isotropic} dynamical models based on HST
photometry and ground based kinematics.

\subsection{A Possible Explanation For the Disappeared Offcenter
    Spike\label{sec:colormap}}

In 1991 the offcenter spike was as strong as the central one which later
flared-up in 1993. It might have been a mere $4\sigma$ statistical
fluctuation, or a nova, a stellar collision, or whatever; we may never
know. But what if a relation exists between the offcenter spike and the
now--confirmed presence of a mini-AGN just $\sim 10$ pc away?

As stated in the Introduction, we know that the surroundings of the
center of NGC~4552 exhibit a quite complex phenomenology, with extended
\ha\ emitting gas (Trinchieri \& di Serego Alighieri 1991) and a dust
ring (Carollo et al.\ 1997a,b).  Relativistic jets are well-known
phenomena associated with AGN, emerging perpendicular to the accretion
disk (e.g. Blandford et al.\ 1991).  It is therefore reasonable to
hypothesize that a jet can be produced each time an accretion event
feeds the supermassive central BH. Such a jet by impacting against a
relatively dense and cold cloud, could shock heat the cloud material,
producing an optically bright transient source.

Figure~\ref{dustring} shows the $V-I$ (F555W-F814W) color map of the
central regions of NGC~4552. The two optical WFPC2 images were obtained
on January 1997 (PI S. Faber). The images are preliminarily reduced by
the standard STScI data pipeline, then coadded using the IRAF/STSDAS
task CRREJ. The resulting color map is resampled onto a finer grid by
means of bilinear interpolation to smooth the pixellated appearance of
the image. The photometric zero point is converted to the
Johnson-Cousins system following the procedures described by Holtzman et
al.\ (1995).

An inhomogeneous ring-like feature clearly emerges with its inner edge
at $\sim 0\farcs1-0\farcs2$ from the center.  This ring is redder than
the surrounding galaxy background with $(V-I)\simeq1.34$, reaching a
peak color difference at P.A.=45$^\circ$ of $\Delta(V-I)\simeq0.10$, or
$E(B-V)\simeq0.09$. The likely interpretation of this feature is that we
are seeing an obscuring dusty ring circling the central spike that is
unobscured as indicated by the  Balmer decrement in the FOS spectra
which is consistent with no absorption.

It is interesting to note that the position of the 1991 offcenter spike
($\sim 0\farcs14$ from the center) places it {\it just} on the inner
edge of the ring, and it is tempting to speculate that the offcenter
spike in the 1991 image could be the result of a transient nuclear jet
impinging on this dusty ring.

Incidentally, it is also interesting to note that the central spike is
clearly seen as a point-like source in Figure~\ref{dustring}. The F555W
and F814W band flux reported by Carollo et al.\ (1997a) for this spike
is consistent with being due uniquely to the line emission (mostly
\ha+\nii) as measured on our FOS spectra, when convolving it with the
transmission curve of the filters and the QE of the detectors.

\subsection{Central Spikes in Other Galaxies}

Pre-COSTAR, deconvolved HST observations of an unbiased sample of early
type galaxies has shown that up to $\sim25\%$ of them may exhibit a
central spike in the visible, in excess of the nuker law (Lauer et al.\
1995). Maoz et al.\ (1996) show that, in FOC images that random sample
the cores of 110 nearby galaxies in the ultraviolet, $\sim10\%$ show a
unresolved point source in their centers.  In the Carollo et al.\
(1997a,b) sample of 18 ellipticals (mostly with kinematically distinct
cores) 4 exhibit a central spike in optical WFPC2 images, one of which
is NGC~4552.

The question that naturally arises is whether any of these other central
spikes are flares caught at some phase in their development.  This is a
non-trivial question, as we know of at least one other galaxy, the Sa
galaxy NGC~2681, which has a UV-bright, unresolved core that is {\it
not} a flare (Cappellari et al.\ 1998).  We see at least two possible
origins for central spikes that are not due to accretion onto a massive
BH:  a recent burst of star formation  or stellar coalescence  in a very
high density environment. Both options are currently being entertained
for the origin of the central cluster of hot stars in the nucleus of our
own Galaxy (Genzel et al.\ 1994), with many such stars exhibiting
Wolf-Rayet type spectra. This central cluster shares two relevant
properties  with some of the spikes mentioned above: a sub-parsec size,
and a luminosity $\sim 10^7\,\lsun$, well in the range covered by them.
Seen from the distance of the Virgo cluster, the bulge of our own Galaxy
would also show an unresolved, UV-bright central spike!

As mentioned in the introduction, a point like source (at the HST
pre-COSTAR resolution) was also detected at the center of the bulge of
the Sa galaxy NGC~2681 (Bertola et al.\ 1995). FOC and FOS observations
similar to those of 1996 reported here for NGC~4552 were also obtained
for this galaxy, and reveal that no excess exists over a pure nuker law
of the power law type (Cappellari et al.\ 1998).

\subsection{Yet Another Way to Search for Central Black Holes in
    Galaxies}

In Section~\ref{sec:low_agn} we have discussed various possibilities for
feeding a central supermassive BH in galactic spheroids. Such masses sit
at the bottom of the gravitational potential well of a galaxy, where
stellar and ISM densities reach their maximum, and where any
cannibalized material tends to converge.  As such, one is tempted to say
that the real problem is how to {\it avoid} a low, fluctuating level of
accretion onto a massive BH --- hence of low level AGN activity ---
rather than how to produce it. We actually argue that wherever a massive
BH exists, it is likely to be accompanied by at least a weak level of
{\it activity}.

The case of NGC~4552 offers a lesson in this respect. Thanks to its
angular resolution, HST observations in either UV or optical imaging or
narrow aperture spectroscopy allow to reveal mini-AGN activity which
would be essentially invisible with similar ground based observations.
Therefore, besides the stellar and gas dynamical techniques mentioned in
the introduction, HST observations similar to those conducted on
NGC~4552 can help to reveal the presence of central BHs in galactic
spheroids.

It is the angular resolution properties of HST that make it so valuable
in the search for nuclear spikes. A similar high resolution capability
is now within reach also from the ground at near-IR wavelength, as a
result of adaptive optics (AO) developments. AO-fed, near-IR, narrow
aperture spectroscopy of the center of galactic spheroids might reveal
broad emission lines of the Paschen and Brackett series, hence
potentially  offering additional opportunities for the demography of
central BHs in galactic spheroids.

This is indeed a subject of great interest in the context of galaxy
formation and evolution. Only few galaxies may indeed host a massive BH
in hierarchical models in which most of star formation takes place in
small galaxies, and it is the late merging of these entities that leads
to the formation of massive spheroids. However, central massive BH may
be more ubiquitous in other hierarchical models in which most of the
merging activity takes place at early times, among mostly gaseous
components, and such merging is accompanied by intense star formation
and dissipational collapse in the central regions. As noted in the
Introduction, central massive BHs seem to be present in most spheroids,
which together with the noted tight relation between the mass of the BH
and the mass of the spheroid seem to argue in favor of the latter
scenario of spheroid formation.

\section{CONCLUSIONS}

By virtue of the normal elliptical galaxy NGC~4552 having been multiply
observed with HST via imaging (in 1991, 1993 and 1996) and via
spectroscopy of its nucleus (1996, 1997), we appear to have caught in
mid-action a transient accretion event onto the central supermassive
black hole of this galaxy.

This accretion reveals itself in several ways: \begin{enumerate}

\item A central, unresolved (at the 0\farcs03 level) ultraviolet-bright
source was first detected in 1991, it increased in luminosity by a
factor of $\sim 4.5$ by 1993, and then declined a factor of $\sim 2.0$
by 1996. Based on its UV spectral energy distribution, we estimate its
1996 luminosity to be $\sim 3\times 10^5\lsun$ and its temperature to be
about 15,000 K.

\item The FOS spectrum of this source taken in May 1996 shows a
Seyfert/LINER type of emission line spectrum, with both permitted {\it
and} forbidden lines being best modelled by the combination of broad
(FWHM $\simeq 3000$ km s$^{-1}$) and narrow (FWHM $\simeq 700$ km
s$^{-1}$) components. Nevertheless, the low luminosity of the broad
H$\alpha$ component ($\sim 5.6\times 10^{37}$ erg s$^{-1}$) makes the
central spike in NGC~4552 likely to be the least luminous among all know
AGNs.

\item The FOS spectrum taken in January 1997 shows quite similar
characteristics, except that the radial velocity of the \ha\ emission
has shifted blueward by $\sim 230$ km s$^{-1}$ compared to the spectrum
taken in May 1996.

\item The 1991 UV image shows a second spike, offset by 0\farcs14 from
the central one, which does not appear in later images.  We speculate
that this second spike could be due to dense gas in the dusty ring seen
at nearly the same distance having been shocked by a relativistic jet
emerging from the central mini-AGN, possibly produced by a previous
accretion event.

\end{enumerate}

From these observations we can constrain the mass of the supermassive
black hole in the center of NGC~4552 to be between $3\times10^8$ to $2
\times 10^9$ \msun, consistent with ground-based estimates. We discuss
various possibilities for the source of matter that can produce an AGN
that is as low in luminosity as this mini-AGN in NGC~4552.  Partial
tidal-stripping of a star in a close fly by with the black hole is an
obvious possibility, but other interpretations cannot be excluded. These
include accretion of clumpy gas that is seen near the center of the
galaxy, or from a mini-inflow gas fed by the X-ray emitting hot gas in
the galaxy. However, the change in the radial velocity of \ha\ over a 8
month period is most consistent with a partial tidal stripping event.

It is likely that most spheroids harbor a central massive black hole, as
suggested by recent observations (cf. Kormendy \& Richstone 1995;
Magorrian et al.\ 1998; Ho 1998). If so, given all the above mentioned
opportunities for a low, fluctuating rate of mass accretion onto such
black holes it appears that transient mini-AGN activity similar to that
we have discovered in NGC~4552 should be a widespread phenomenon.
Indeed, it is quite conceivable that each of all mentioned mechanisms --
tydal stripping, accretion from a clumpy ISM, accretion from a
mini-inflow -- are occasionally at work in real galaxies. While order of
magnitude estimates of the duration and frequency of such events might
be guessed from a theoretical viewpoint, we believe that first expanding
the empirical evidence will be more rewarding in the near future.

HST high resolution UV imaging and narrow aperture spectroscopy with
either HST of with adaptive optics from the ground should allow to
detect in other galaxies signs of similar mini-AGN activity.  This will
offer yet another opportunity to expand the demography of central
massive black holes in galactic spheroids, as well as to study virtually
unobscured black hole accretion phenomena at a different regime compared
to more powerful AGNs.

\acknowledgements

MC is grateful to the European Southern Observatory for its kind
ospitality in the period during which much of the HST data analysis has
been completed. DB acknowledges support by NASA/STscI through grants
GO-03728.01-91A and GO-06309.01-94A to DB. LG is grateful to the
Observatory of the University of Munich for its extensive hospitality
and for the providing a most stimulating environment.


\appendix

\section{Nonlinearity Correction of a  Non-Uniform
Background\label{sec:nonlin}}

Flat field nonlinearity of the FOC has been known for a long time and
has been well-characterized (Jedrzejevski 1992). The true count rate
$\rho$ can be derived from the observed count-rate $r$ by inversion of
the following equation, which describes the behavior of the linearity
relation for intensity values up to $\sim80\%$ of the saturation value
$a$
\begin{equation}
    r = a\ (1-e^{-\rho / a}) \label{eq:nonlin}
\end{equation}
where $a$ is a fitting parameter that has been determined from actual
measurements. $a=0.73$ for the 512$\times$512 F/96 observing mode and
$a=0.11$ for the 512z$\times$1024 mode (de-zoomed pixel!) (Nota et al.\
1996).

It has also long been known that point sources remain in the FOC linear
range up to values $\sim8\times$ higher than those for the flat field.
This has lead for example in the FOC Handbook (Nota et al.\ 1996) to the
simple guideline for observer to ``keep the count rate in the central
pixel below 1 count/sec,'' to get reliable photometry of point sources
in the 512$\times$512 mode.

It was not so clear what to do in the intermediate cases of a
non-uniform background, as is always the case with the galaxies surface
brightness photometry. In some  cases the above guideline was still
adopted, leading to severe underestimation of the nonlinearity effect
(eg. Crane et al.\ 1993), as will be shown below.

Greenfield (1994) presented a simple method to correct for nonlinear
behavior of stars over a substantial background count rate. This method
requires measuring the flux of the point source in an appropriate
aperture of radius $r_n$ and consists of applying the flat field
nonlinearity correction of Equation~\ref{eq:nonlin} to the average flux
in that aperture. The appropriate radius to be used with the
512$\times$512 format is $r_n=5.5$ pixel, while for the 512z$\times$1024
it is $r_n=8.6$ pixel (applying the correction to the de-zoomed image).

In this paper we extend the above method to deal with more general
non-uniform objects, in a way that was already suggested, but not tested
by Greenfield. This method consists of smoothing the nonlinear image
using a simple circular smoothing aperture having the radius $r_n$
determined by Greenfield. The correction factor is then computed for
each pixel of the smoothed image and finally the correction is applied
to the corresponding pixel of the original image. We have tested this
method using it to correct our severely nonlinear (maximum correction
factor $\sim1.8$) image of NGC~4552 obtained in 1993 with the F342W
filter and comparing this photometric  profile with that of the 1991
image, again corrected with the above algorithm (even though the
nonlinearity does not exceed $\sim5\%$). We have found that the two
profiles agree to better than 2\% over the whole intensity range.

The inverse of the above method will be used in this paper to simulate
the effect of the FOC nonlinearity over a synthetic image of a galaxy,
which will represent the ``true'' galaxy image.  In this case one has to
smooth the synthetic image, again  using a circular smoothing aperture
of radius $r_n$. The nonlinearity of each pixel is computed from the
smoothed version of the synthetic image using Equation~\ref{eq:nonlin}
and the nonlinearity is finally applied to original synthetic image.

As can be seen in Figure~\ref{F342W_1991_NGC4552} this method of
simulating the nonlinearity effect is able to reproduce very well
(within 0.05 mag from $0\farcs1$ to $3''$ from the center of the galaxy)
the sharp decrease of the 1993 F342W profile [panel (d)], compared to
the 1991 F342W one [panel (g)]. In both cases the starting synthetic
galaxy is exactly the same apart from a scaling factor.



\clearpage
\epsscale{.4}
\plotone{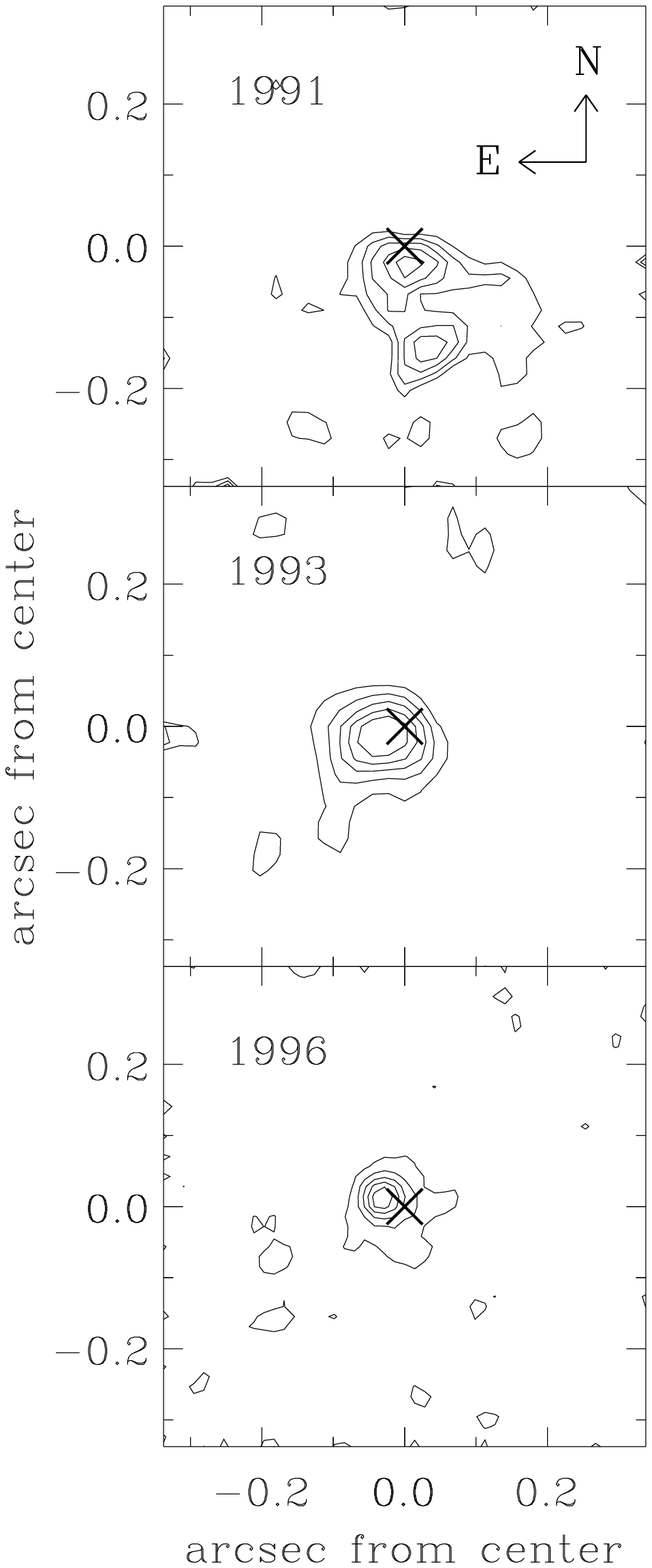}
\figcaption{Contour plots of the very central region of NGC~4552 as
recorded by FOC in the F342W waveband (FoV=$0\farcs6\times0\farcs$6).
North is up and East to the left. Top, middle and bottom panel
correspond to the 1991, 1993 and 1996 observations, respectively. The
cross represents the photometric center of the galaxy as obtained by
fitting to the light distribution of the galaxy an elliptical annulus
with semi-major axis $a=1''$ and width $\Delta a=1''$. At all the three
epochs the position of the spike coincides with the center of the galaxy
within the errors ($\sim0\farcs03$).\label{Identificazione_Spike}}

\clearpage
\vspace*{3truecm}
\epsscale{1}
\plotone{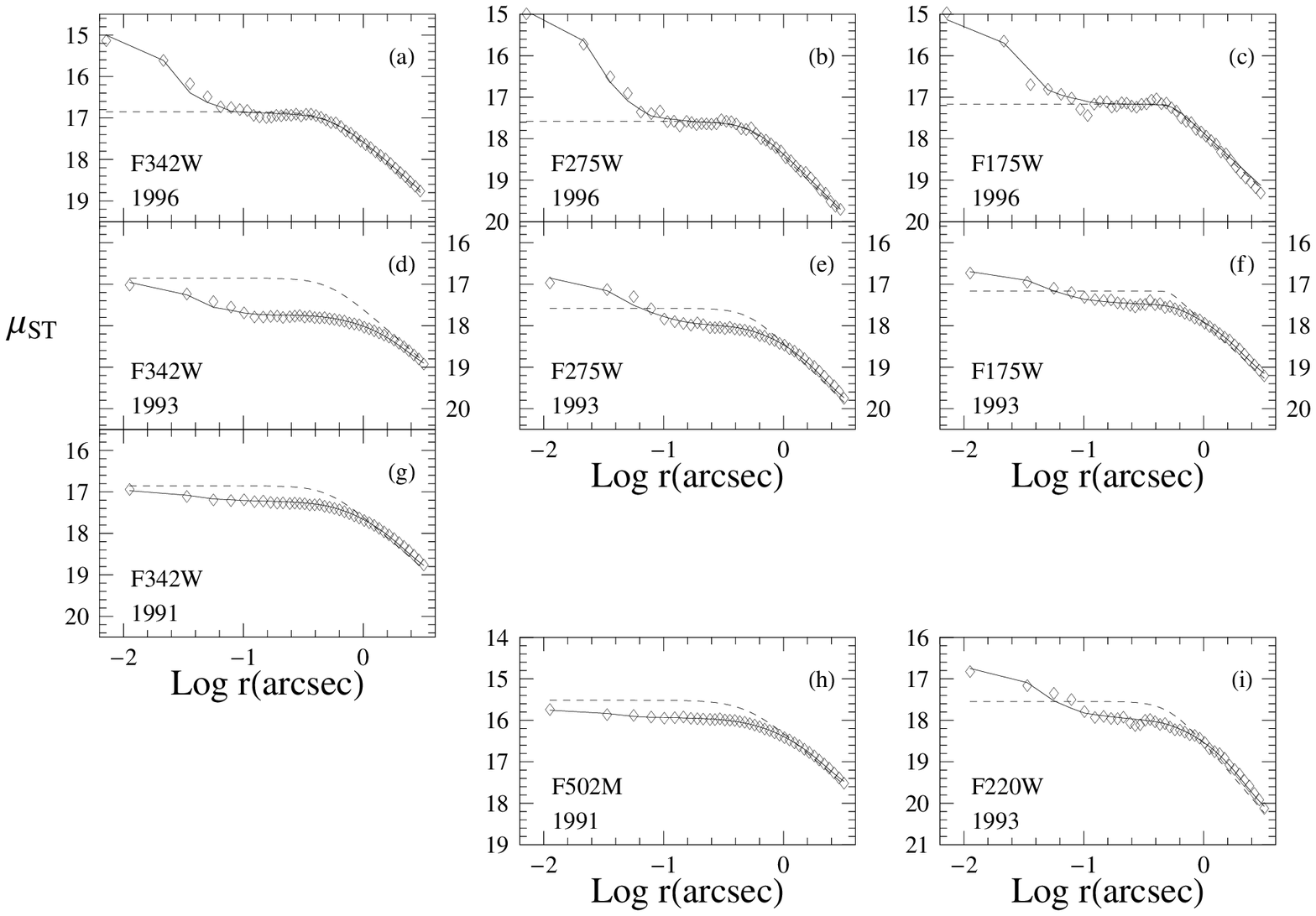}
\figcaption{Modeling of the inner surface brightness profiles (HST
magnitudes) of NGC~4552 vs. $\log r$ in different FOC wavebands.
Diamonds represent the observed profile, the dashed line represents a
model of the true galaxy profile and finally the solid line shows the
above model after adding a point-like central source and then applying
all the known instrumental effects of that specific observation (see
text for details). Note that, within a specific waveband, the underlying
(spike-free) galaxy model has been constrained to be the same at all
epochs.\label{F342W_1991_NGC4552}}

\clearpage
\vspace*{3truecm}
\epsscale{.6}
\plotone{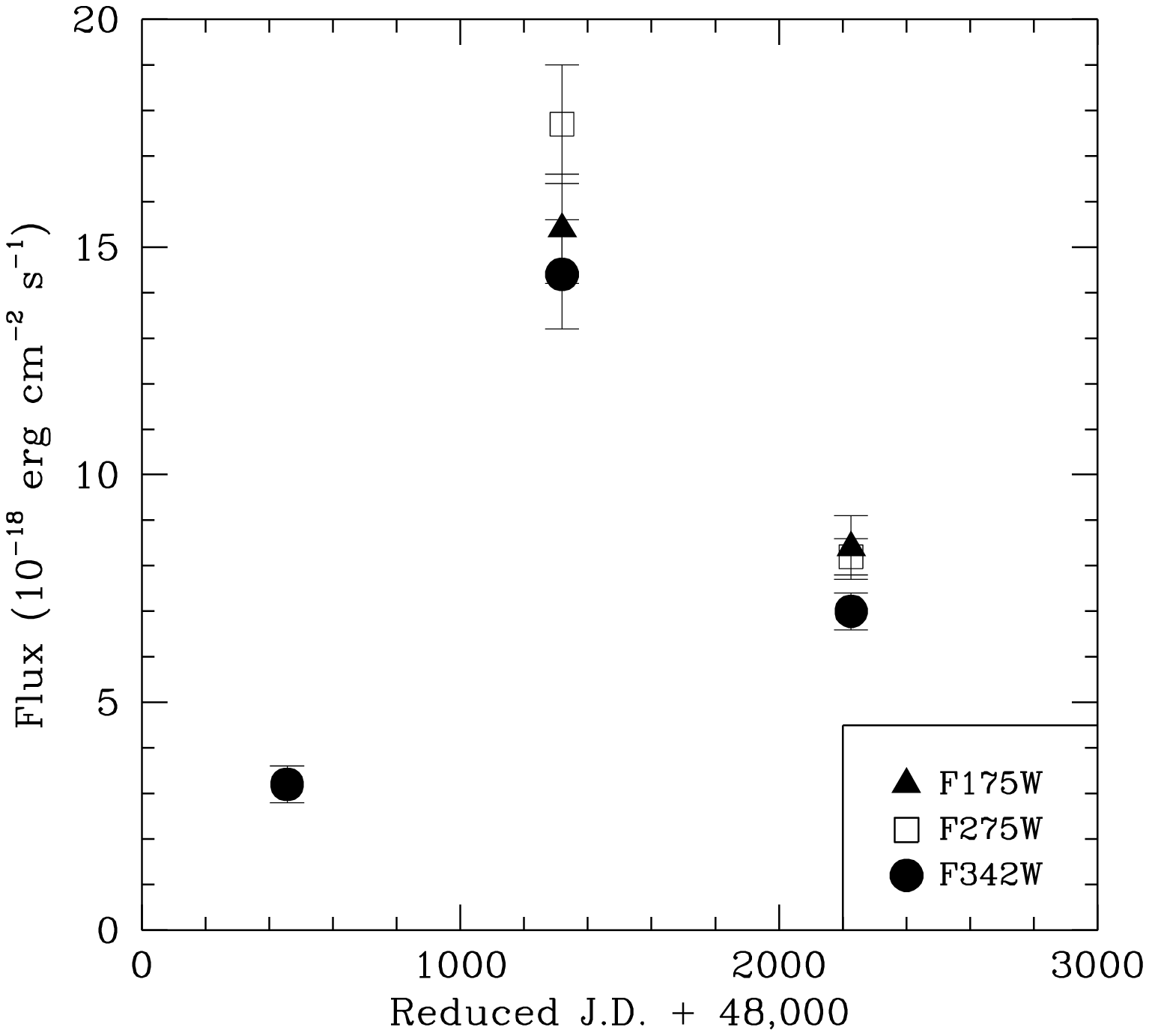}
\figcaption{The light curve of the spike in the  F175W, F275W and F342W
passbands.\label{variazione_spike}}

\clearpage
\vspace*{3truecm}
\epsscale{.7}
\plotone{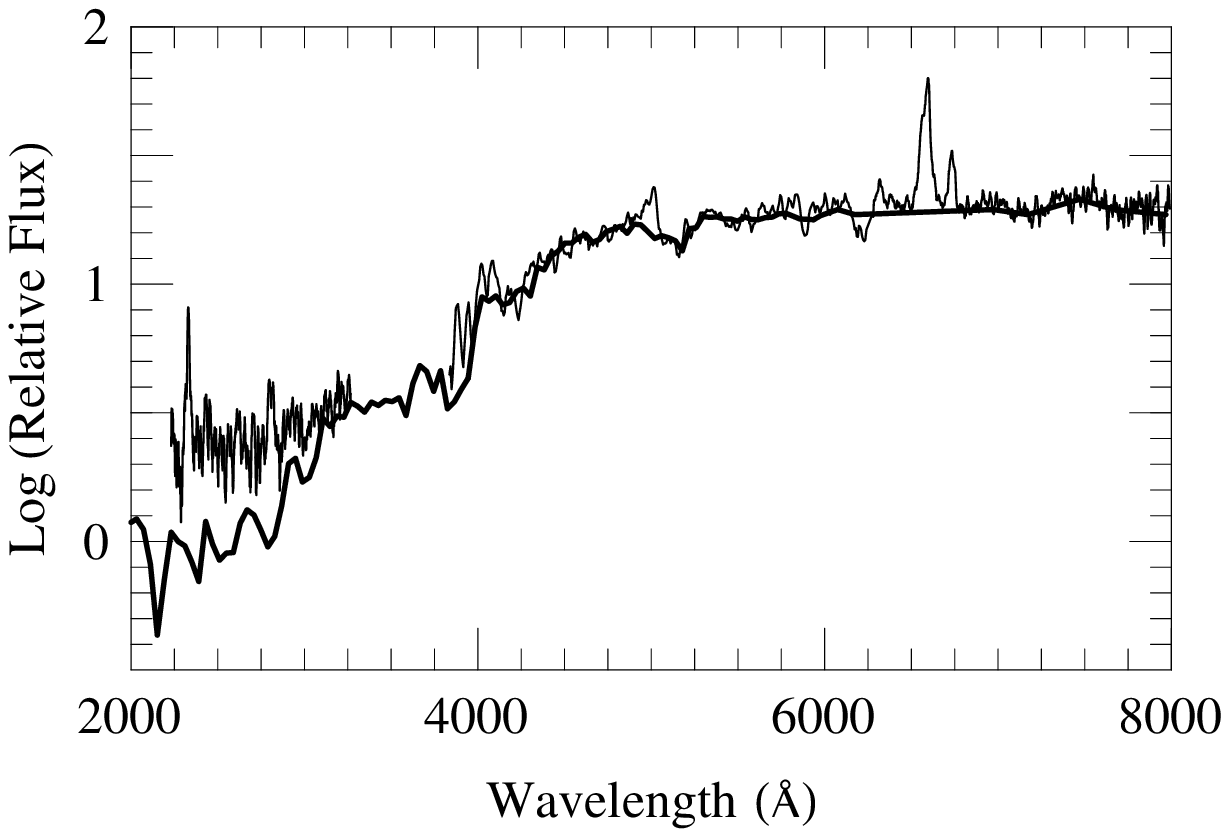}
\figcaption{The overall 1996 FOS spectrum of NGC~4552 within the
$0\farcs2\times0\farcs2$ aperture centered on the spike (thin  line), is
superimposed to a scaled combination of the IUE
10\arcsec$\times$20\arcsec\ aperture of NGC~4552 (Burstein et al.\ 1988)
matched to ground-based optical spectrum of NGC~4649, a giant elliptical
whose SED is virtually the same as that of NGC~4552 (Oke et al.\ 1981;
thick line).  The spectra have been normalized to the visual region. The
FOS spectrum appears quite different owing to the appearance of UV and
optical emission lines as well as a continuum UV excess shortward of
$\lambda\sim3000$~\AA.\label{Spettro_Completo_Spike_N4552}}

\clearpage
\vspace*{3truecm}
\epsscale{.7}
\plotone{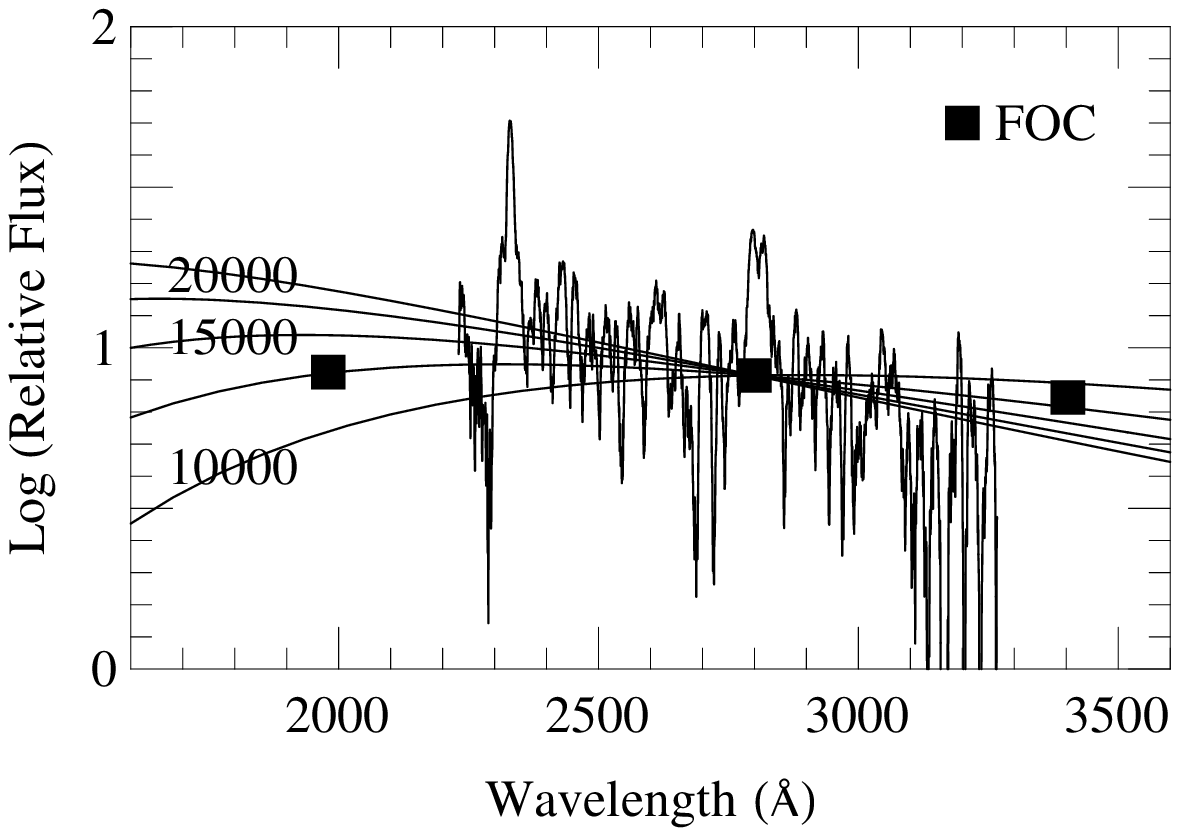}
\figcaption{Comparison between the starlight-subtracted UV (G270H)
portion of the 1996 FOS spectrum  of the spike region and our FOC
(F175W, F275W, F342W) UV photometry. The subtracted SED corresponds to
the IUE spectrum shown in Fig.~\ref{Spettro_Completo_Spike_N4552}. The
best match  of both FOS spectrum and FOC photometry (filled squares)
with the superimposed blackbody energy distributions is for
$T\sim15,000$~K. Moreover, such a comparison indicates that the UV
continuum comes essentially from the unresolved
spike.\label{Continuo_Spike_N4552}}

\clearpage
\epsscale{.7}
\plotone{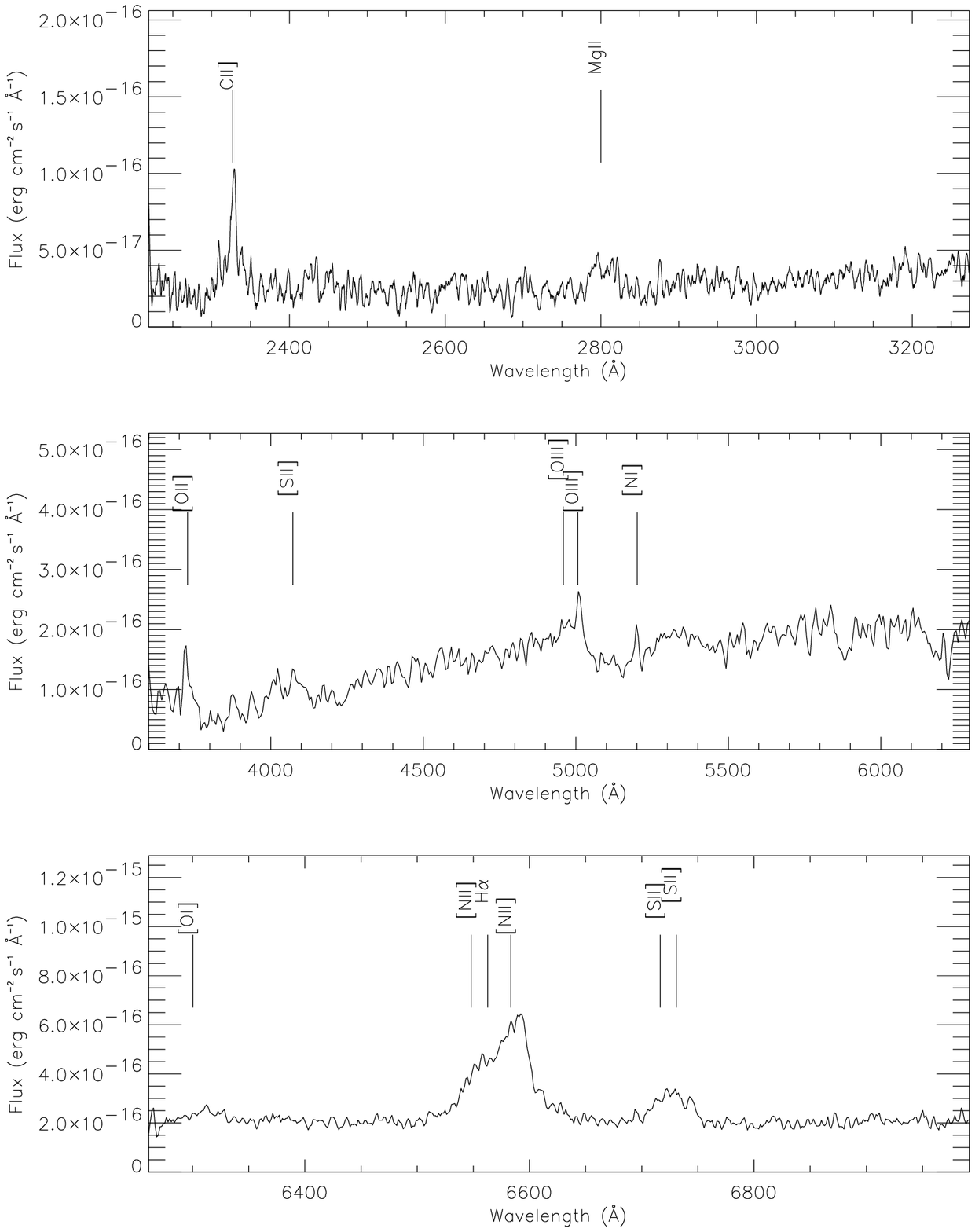}
\figcaption{The 1996 HST/FOS spectra of the central
0\farcs2$\times$0\farcs2 region of NGC~4552 . Each panel shows the
spectrum from a different grating, in order of increasing wavelength
(G270H, G650L and G780H). The spectra are slightly boxcar smoothed.
Identification of the most prominent emission lines are
given.\label{Identificazione_Righe_N4552}}

\clearpage
\epsscale{.7}
\plotone{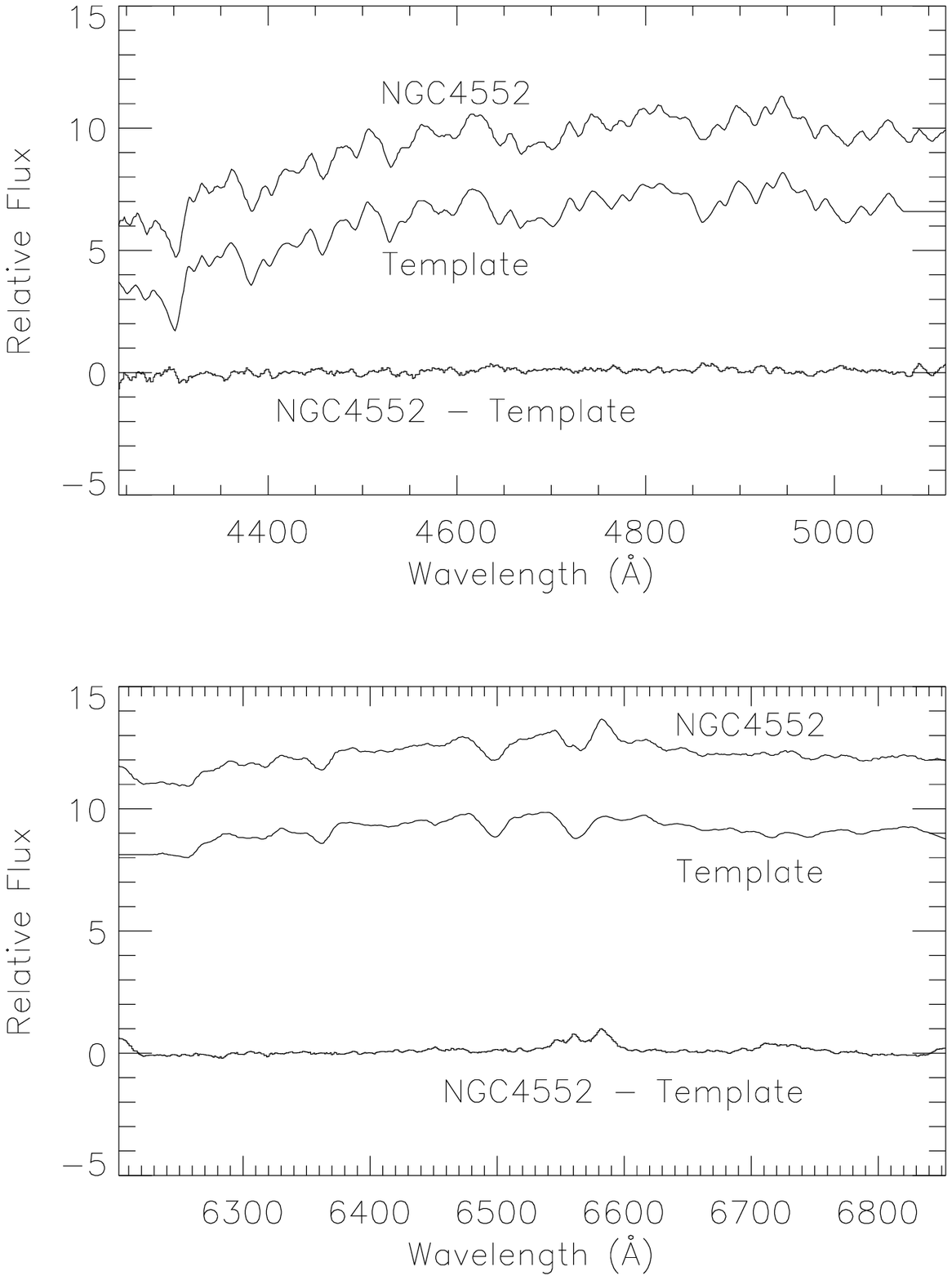}
\figcaption{Underlying starlight subtraction from the spectrum of
NGC~4552 by means of the adopted emission-free  template spectrum
(NGC~3115). Both spectra are from the library of Ho et al.\ (1995) and
refer to an aperture of 2\arcsec$\times$4\arcsec. Upper panel: the
observed blue portion of the spectrum (top plot) is compared with the
template (middle plot, offset by a constant), while the bottom plot
shows the difference between the spectrum of NGC~4552 and the template.
Lower panel: the same as in upper panel for the red portion of the
spectrum.  The excellent match in the spectral region free from emission
lines fully justifies the choice of NGC~3115 as a template for starlight
subtraction within the FOS aperture (note that a resolved \ha+\nii\
emission emerges for NGC~4552).\label{template_subtraction}}

\clearpage
\vspace*{3truecm}
\epsscale{1}
\plotone{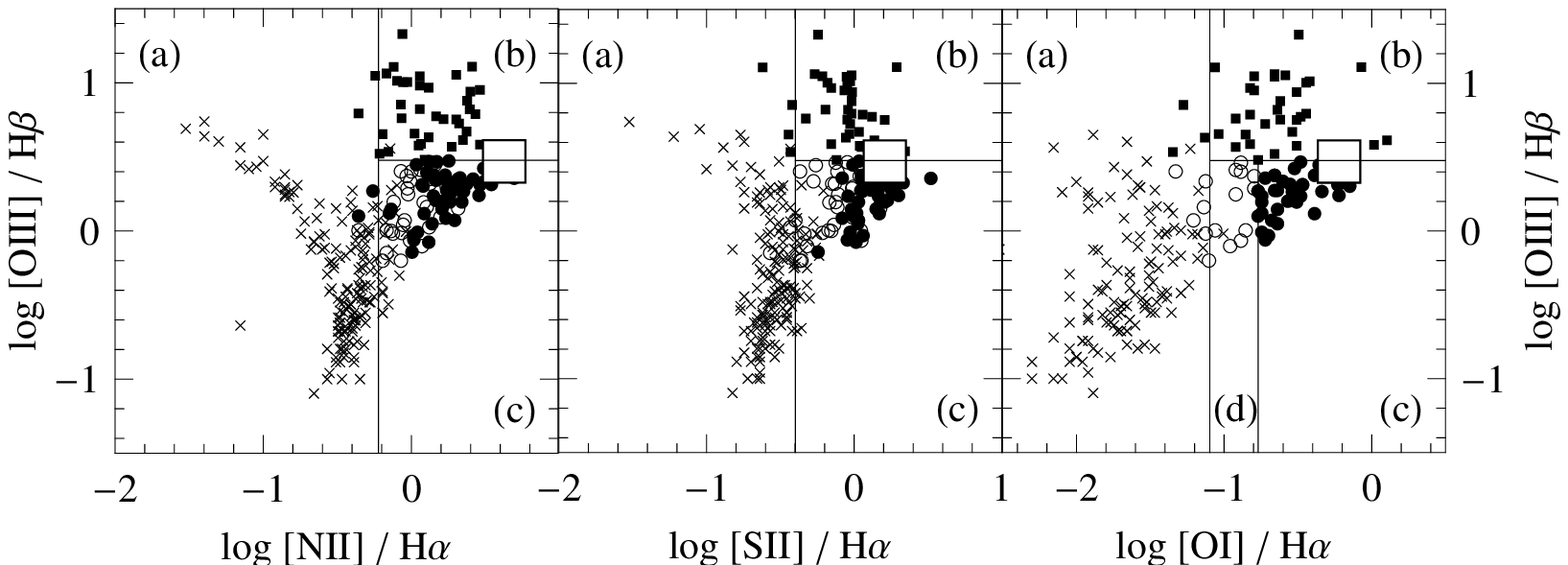}
\figcaption{The location of the NGC~4552 nucleus (as derived from the
narrow line emission components measured on the FOS spectra) on the
diagnostic diagrams used by Ho et al.\ (1997); the large open square
represents the position of NGC~4552 at the 1997 epoch. The corresponding
errors are of the size of the smaller simbols. The 1996 data are
consistent with the 1997 measurements, although with larger errors. The
other symbols represent the nuclei included in the Ho et al. sample
(crosses = \hii\ nuclei, filled squares = Seyfert nuclei, filled circles
= LINERs, open circles = transition objects). The vertical and
horizontal lines delineates the boundary adopted by Ho et al.\ between
(a) \hii\ nuclei, (b) Seyfert galaxies, (c) LINERS and (d) Transition
objects. \label{classificazione_spike_n4552}}

\clearpage
\epsscale{.4}
\plotone{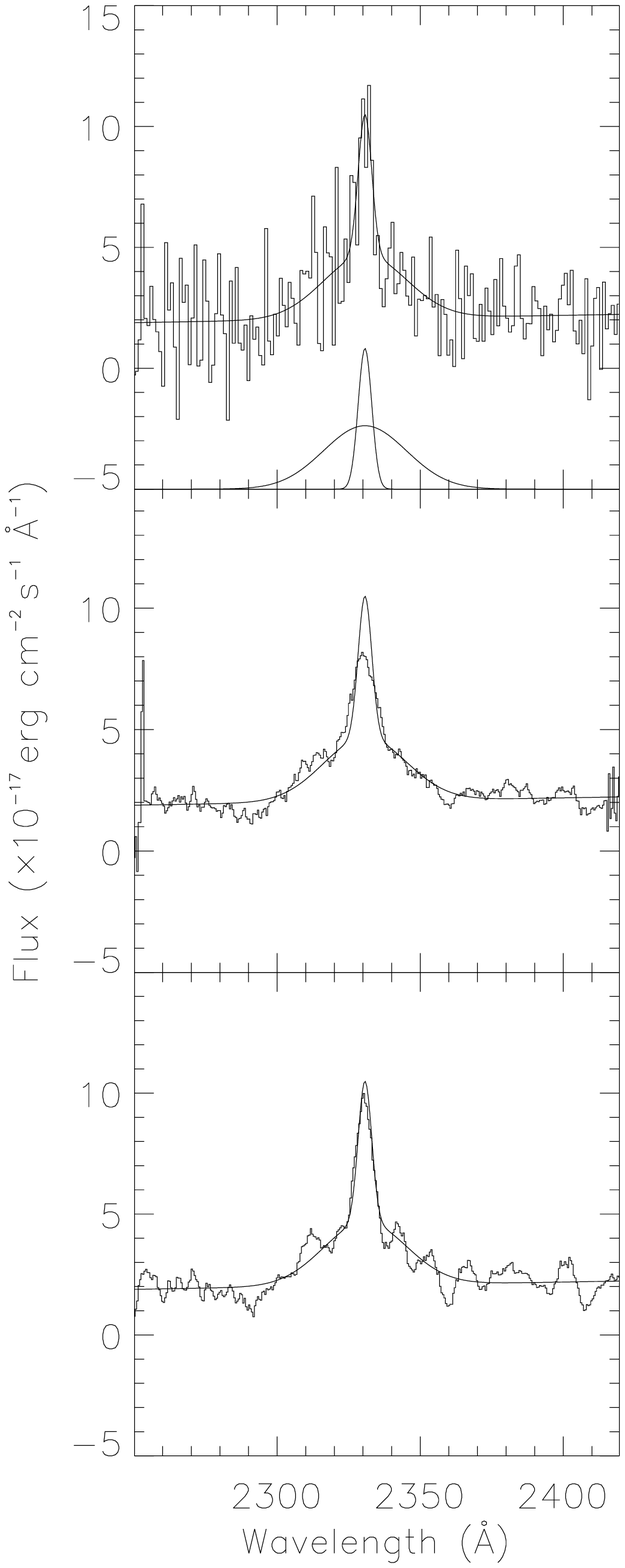}
\figcaption{Upper panel: The portion of original UV (G270H) spectrum
containing the \cii~$\lambda$2326~\AA\ multiplet with superimposed our
best-fit model; individual components of the model are shown at the
bottom of the panel. Middle panel: the best-fit model is overplotted on
a 19 pixel, boxcar-smoothed version of the spectrum; Bottom panel: the
same best-fit model is superimposed to a 37 pixel, 4th degree
Sawizki-Golay filtered spectrum. The adopted filterings applied to  the
original spectrum as shown in the two lower panels are meant to
emphasize the broad and narrow component, respectively. The reduced
chi-square for the fit in the upper panel is
$\chi^2_\nu\simeq1.06$.\label{modeling_cii}}

\clearpage
\epsscale{.65}
\plotone{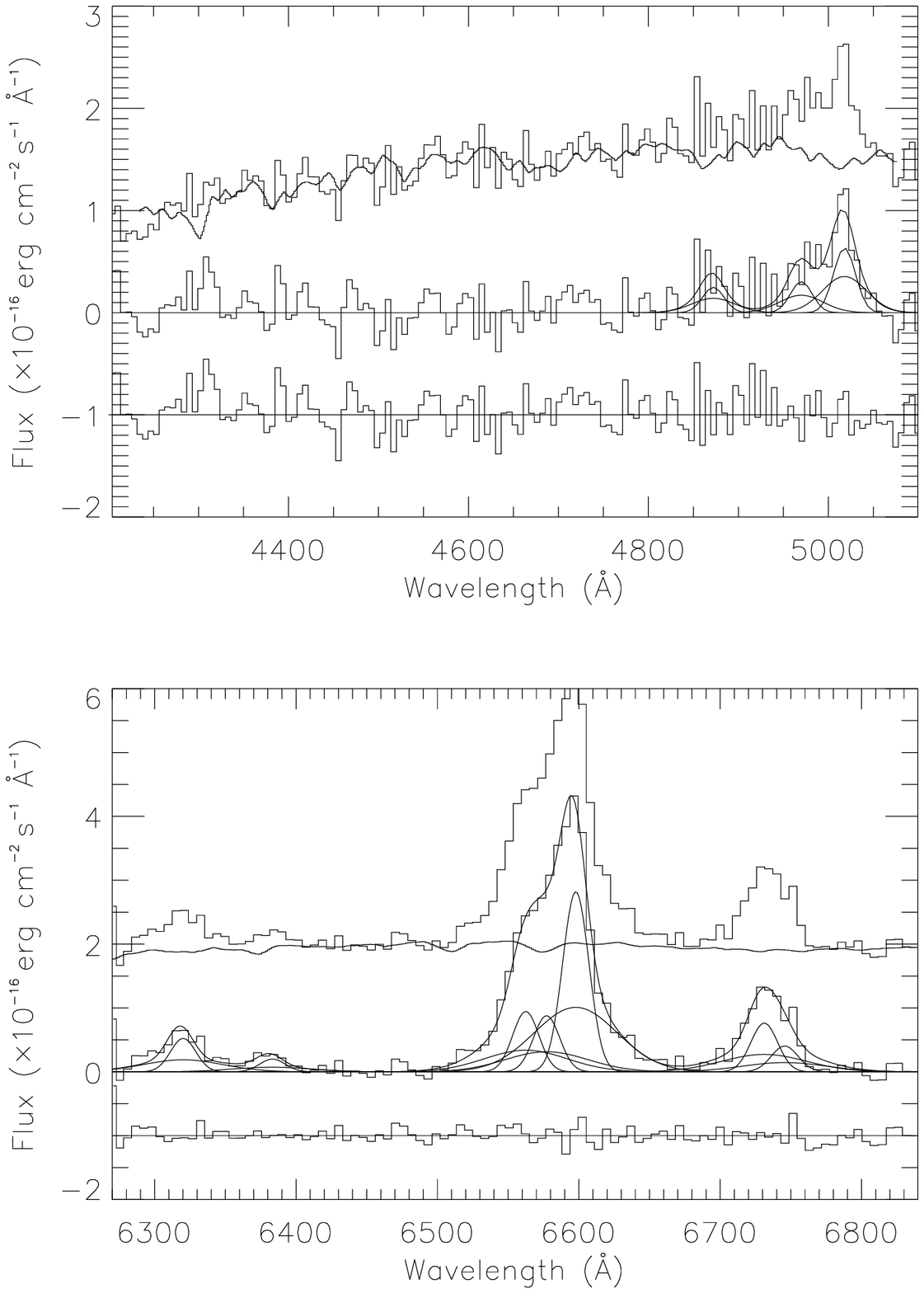}
\figcaption{Upper panel: gaussian decomposition of the H$\beta$ and
\oiii\  lines from the  FOS G650L spectrum of the center of NGC~4552. In
the top plot the adopted emission-free template is superimposed to the
original spectrum. The middle plot represents the starlight-subtracted
spectrum with superimposed our best-fit Gaussuan model as well as the
individual (narrow and broad) components. The residuals (arbitrarily
shifted) are shown at the bottom of the panel. Lower panel: the same as
in upper panel for the \oi, \nii, \ha, and \sii\ lines in the red region
of the FOS G780H spectrum, rebinned over 4 pixels. Owing to the low S/N
of the G650L spectrum the lines included in in this range have been
modeled imposing the same gaussian parameters obtained from the lines in
the  G780H spectrum, hence determining only redshift and line fluxes.
The reduced chi-square for the fit in the upper panel is
$\chi^2_\nu\simeq1.23$, while for the lower panel fit
$\chi^2_\nu\simeq1.00$. \label{modeling_halpha_agn_like}}

\clearpage
\epsscale{.65}
\plotone{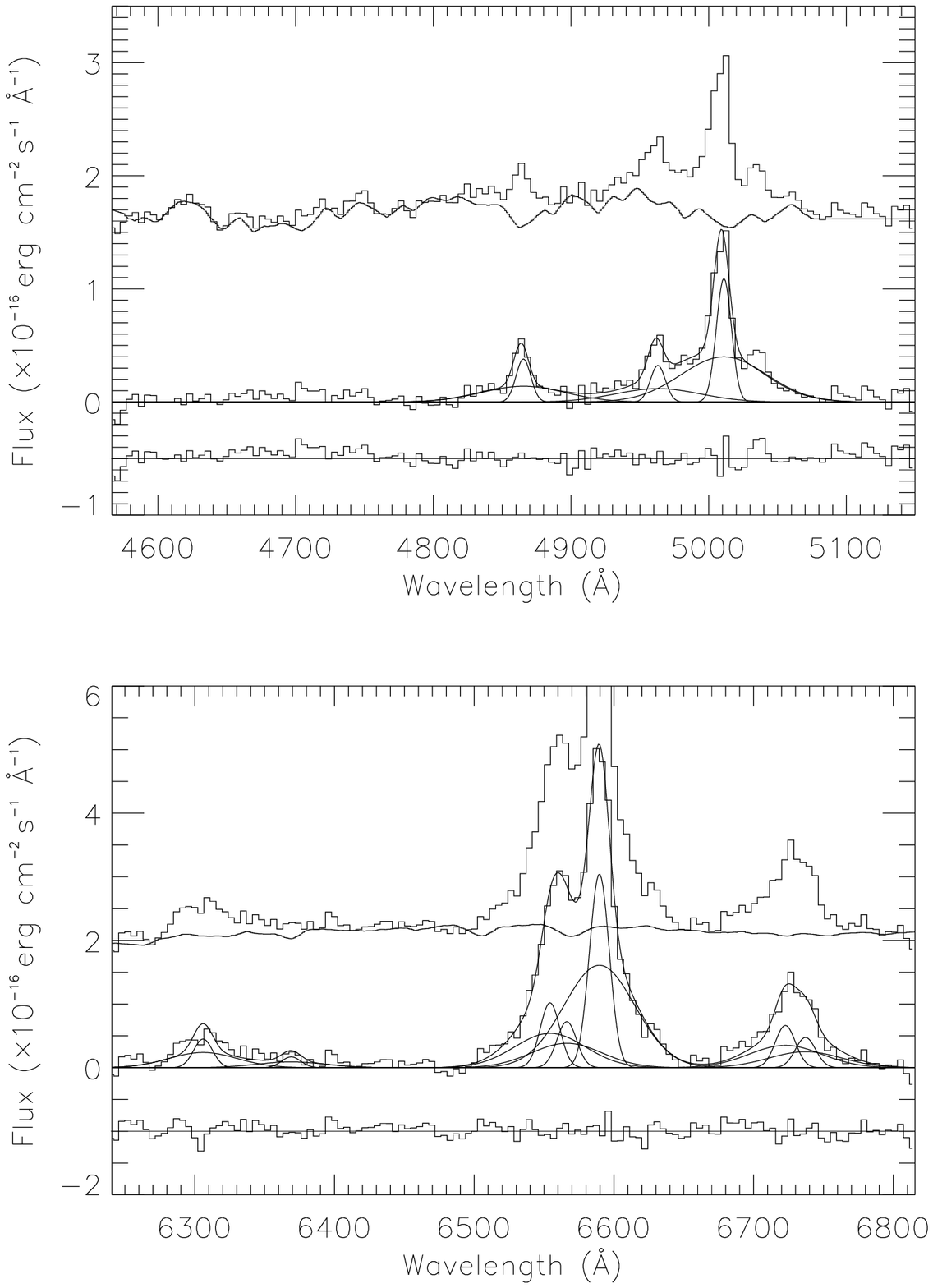}
\figcaption{The same as in Fig.~\ref{modeling_halpha_agn_like} for the
FOS G570H spectrum (here rebinned over 4 pixels) obtained by Faber and
collaborators in January 1997. In order to get a satisfactory fit of the
\ha+\nii\ complex, a blueshift by 4.6~\AA\ ($\sim 230$ km s$^{-1}$) of
both components of \ha\ was required with respect to both 1996 and to
the 1997 position of the \nii\ doublet. For the fit in the upper panel
$\chi^2_\nu\simeq1.35$, while in the lower panel
$\chi^2_\nu\simeq1.37$.\label{modelling_97}}

\clearpage
\vspace*{3truecm}
\epsscale{.5}
\plotone{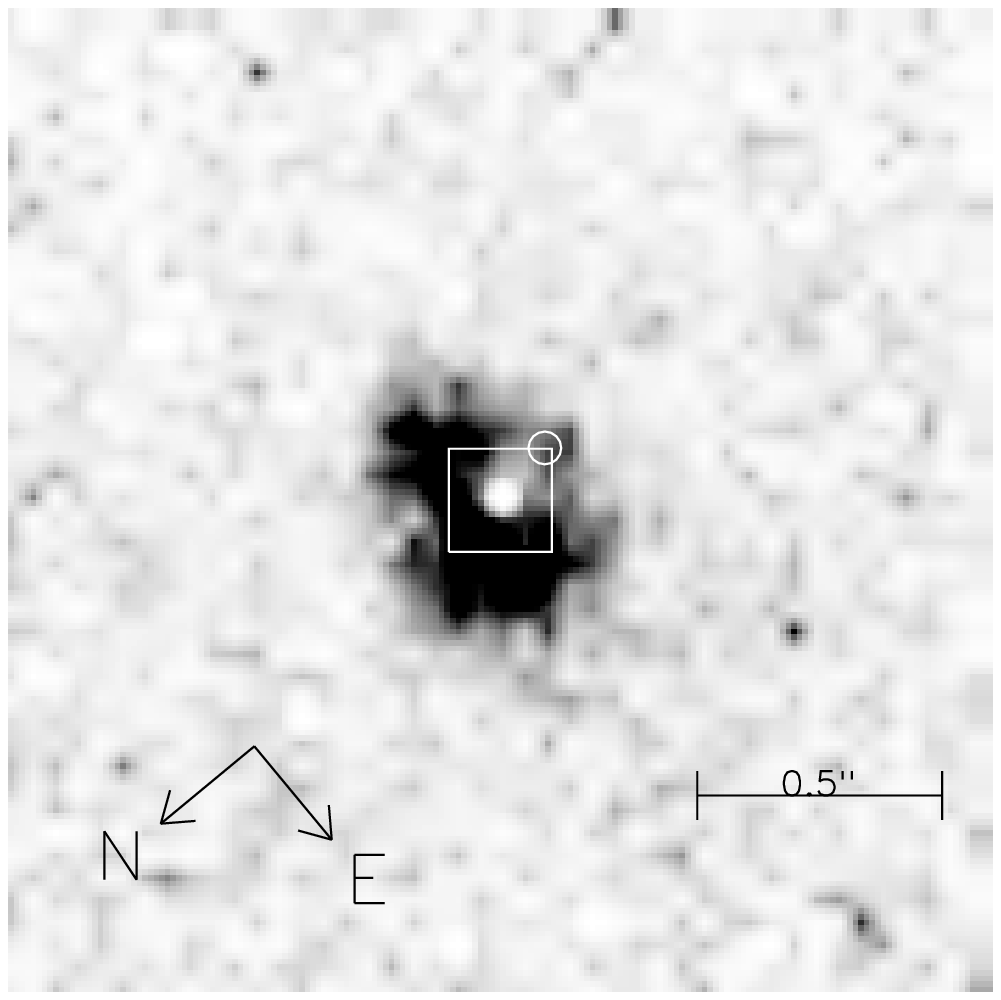}
\figcaption{The $V-I$ (F555W-F814W) color map of the central regions of
NGC~4552. The two optical WFPC2 images were obtained on Jan. 1997. The
darkest color correspond to $V-I\simeq1.44$ mag, while the background is
at the $V-I\simeq1.34$ mag level. The central square represents the size
of the 0\farcs2$\times$0\farcs2 FOS aperture used for the spectroscopic
observations. The central spike is clearly visible as a bright spot, and
the position of the offcenter spike present only in the 1991 data is
indicated by the open circle. \label{dustring}}


\clearpage

\begin{deluxetable}{ccccr}
\tablecaption{FOC f/96 Observation Log. \label{tab:obs_log}}
\tablehead{
\colhead{COSTAR} & \colhead{Observing Mode} & \colhead{Filter} &
\colhead{Date}   & \colhead{Exp.}\\
 &  &  &  & \colhead{(s)}
}
\startdata
    No  & 512$\times$512   & F342W  & 19 Jul 1991 & 1196 \nl
    No  & 512$\times$512   & F502M  & 19 Jul 1991 & 896 \nl
\tableline
    No  & 512z$\times$1024 & F175W  & 28 Nov 1993 &
2453\tablenotemark{a} \nl
    No  & 512z$\times$1024 & F220W  & 28 Nov 1993 & 1377 \nl
    No  & 512z$\times$1024 & F275W  & 28 Nov 1993 & 717 \nl
    No  & 512z$\times$1024 & F342W  & 28 Nov 1993 & 597 \nl
\tableline
    Yes & 512$\times$512   & F175W  & 23 May 1996 & 1287 \nl
    Yes & 512$\times$512   & F275W  & 23 May 1996 & 384 \nl
    Yes & 512$\times$512   & F342W  & 23 May 1996 & 296 \nl
\enddata
\tablenotetext{a}{Sum of two equal exposures}
\end{deluxetable}


\begin{deluxetable}{cccccccc}
\scriptsize
\tablecaption{Fitted Parameters in Equation~\ref{eq:nuker} obtained from
    the Photometric Modeling. \label{tab:phot}}
\tablehead{
\colhead{Filter} & \colhead{Date} &
\colhead{$\alpha$} & \colhead{$\beta$} & \colhead{$\gamma$} &
\colhead{$r_b$} & \colhead{$I_b$} & \colhead{$c_s$} \\
\colhead{(1)} & \colhead{(2)} & \colhead{(3)} & \colhead{(4)} &
\colhead{(5)} & \colhead{(6)} & \colhead{(7)} & \colhead{(8)}
}
\startdata
 F502M  & 1991 & 2.89$\pm$0.34 & 1.020$\pm$0.034 & 0.00$\pm$0.03 &
0.507$\pm$0.022  & 140.1$\pm$4.0 & 1100$\pm$110  \nl
\tableline
 F342W  & 1991 & 3.51$\pm$0.81  & 1.016$\pm$0.061   & 0.00$\pm$0.03 &
0.514$\pm$0.037   & 134.86$\pm$0.70 & 1600$\pm$200 \nl
 ''     & 1993 & ''             & ''                & ''            &
''                & 85.60$\pm$0.89 & 4530$\pm$290 \nl
 ''     & 1996 & ''             & ''                & ''            &
''                & 9.39$\pm$0.44 &  589$\pm$31 \nl
\tableline
 F275W  & 1993 & 4.06$\pm$0.90  & 1.137$\pm$0.055   & 0.00$\pm$0.03 &
0.511$\pm$0.030   & 14.82$\pm$0.15 &  1870$\pm$110  \nl
 ''     & 1996 & ''             & ''                & ''            &
''                & 1.866$\pm$0.082 &  268$\pm$12  \nl
\tableline
 F220W  & 1993 & ''             & 1.35$\pm$0.19\tablenotemark{a}    &
''            & ''                & 11.08$\pm$0.14 &  1165$\pm$84 \nl
\tableline
 F175W  & 1993 & 17$\pm$10      &  1.062$\pm$0.050  & 0.00$\pm$0.03 &
0.517$\pm$0.029   & 14.32$\pm$0.09 &  919$\pm$47 \nl
 ''     & 1996 & ''             & ''                & ''            &
''                & 1.640$\pm$0.089 &  142$\pm$11 \nl
\enddata
\tablenotetext{a}{In the FOC F220W the S/N is low and the Nuker-law
    parameters can not be reliably determined: they have been imposed to
    be equal to those obtained for the F275W frames. On the other hand
    the outer profile appears significantly different from that of the
    other images, even taking into account the large error in the
    background subtraction: for this reason it has been independently
    measured.}
\tablecomments{Col.~(1): FOC filter used in the observation; Col.~(2):
    Date the observation was taken; Col.~(3): The $\alpha$ Nuker-law
    parameter measures the sharpness of the break between the inner and
    the outer power-law profiles; Col.~(4): The $\beta$ parameter is
    the stepness of the outer profile ($I(r) \propto r^{-\beta}$ for $r
    \gg r_{\rm b}$); Col.~(5): The $\gamma$ parameter is the stepness
    of the inner profile ($I(r) \propto r^{-\gamma}$ for $r \ll r_{\rm
    b}$); Col.~(6): $r_b$ is the radius of the break, measured in
    arcsec; Col.~(7): $I_b$ is a scaling factor of the profile and
    indicates the surface brightness at $r=r_b$ in raw units (counts
    pixel$^{-1}$); Col.~(8): $c_s$ are the raw {\em total} counts of the
    spike. See text for details of how they have been measured. A '' in
    this table means that this parameter has been held fixed at the
    value determined in above line, and it has not been fitted. See text
    for details.}
\end{deluxetable}


\begin{deluxetable}{ccccc}
\tablecaption{Variation of the Spike and Relative Calibration of the
    Frames. \label{tab:calib}}
\tablehead{
 & \multicolumn{2}{c}{Spike Variation}
 & \multicolumn{2}{c}{Sensitivity Variation} \\
\cline{2-3} \cline{4-5} \\
\colhead{Filter} & \colhead{$f_{93}/f_{91}$} & \colhead{$f_{93}/f_{96}$}
&
\colhead{$U_{93}^{-1}/U_{91}^{-1}$}  &
\colhead{$U_{93}^{-1}/U_{96}^{-1}$}
}
\startdata
 F175W & \nodata     & 1.8$\pm$0.2 & \nodata       & 1.86$\pm$0.10\nl
 F275W & \nodata     & 2.2$\pm$0.2 & \nodata       & 1.73$\pm$0.08 \nl
 F342W & 4.5$\pm$0.6 & 2.1$\pm$0.2 & 1.27$\pm$0.02 & 1.84$\pm$0.09 \nl
\enddata
\end{deluxetable}


\begin{deluxetable}{cccccc}
\footnotesize
\tablecaption{Calibrated Flux from the Spike. \label{tab:fluxes}}
\tablehead{
\colhead{Filter}   & \colhead{$\lambda_{ eff}$} & \colhead{$f_{91}$} &
\colhead{$f_{93}$} & \colhead{$f_{96}$} \\
\colhead{(1)}  & \colhead{(2)}  & \colhead{(3)} &
\colhead{(4)} & \colhead{(5)}
}
\startdata
 F175W & 1970 & \nodata     & 15.4$\pm$1.2 & 8.4$\pm$0.7 \nl
 F220W & 2320 & \nodata     & 14.8$\pm$1.2 & \nodata     \nl
 F275W & 2800 & \nodata     & 17.7$\pm$1.3 & 8.2$\pm$0.4 \nl
 F342W & 3400 & 3.2$\pm$0.4 & 14.4$\pm$1.2 & 7.0$\pm$0.4 \nl
 F502M & 4990 & 7.0$\pm$0.8 & \nodata & \nodata \nl
\enddata
\tablecomments{Col.~(1): FOC filter used in the observation; Col.~(2):
Effective wavelength of the filter in \AA; Col.~(3-5): Total flux from
the spike in units of $10^{-18}$ \esca.}
\end{deluxetable}


\begin{deluxetable}{ccccc}
\tablecaption{FOS Observation Log. \label{tab:fos_log}}
\tablehead{
\colhead{Grating} & \colhead{$\lambda$ range} &
\colhead{Spectral Resolution} & \colhead{Date} & \colhead{Exp.} \\
    & \colhead{(\AA)} & \colhead{(FWHM \AA)} & & \colhead{(s)}
}
\startdata
 G270H & 2222--3277 & 1.89 & 24 May 1996 & 1240 \nl
 G270H & 2222--3277 & 1.89 & 24 May 1996 & 1110 \nl
 G650L & 3540--7075 & 23.4 & 24 May 1996 &  400 \nl
 G780H & 6270--8500 & 5.26 & 24 May 1996 &  430 \nl
 G780H & 6270--8500 & 5.26 & 24 May 1996 & 2410 \nl
\tableline
 G570H & 4569--6818 & 4.02 & 16 Jan 1997 & 1380 \nl
 G570H & 4569--6818 & 4.02 & 16 Jan 1997 & 2410 \nl
 G570H & 4569--6818 & 4.02 & 16 Jan 1997 & 2410 \nl
 G570H & 4569--6818 & 4.02 & 17 Jan 1997 & 2410 \nl
 G570H & 4569--6818 & 4.02 & 17 Jan 1997 & 1690 \nl
\enddata
\end{deluxetable}


\begin{deluxetable}{cclcccccccccc}
\scriptsize
\tablecaption{Line Emission Fluxes and Modeling
Parameters\label{tab:line_modeling}}
\tablehead{
\colhead{Spectrum} & \colhead{Date} & \colhead{Line} & \colhead{Line
Flux NLR} &
\colhead{BLR/NLR} & \colhead{FWHM$_{NLR}$} &
\colhead{FWHM$_{BLR}$} & \colhead{LOS v}\\
\colhead{(1)} & \colhead{(2)} & \colhead{(3)} & \colhead{(4)} &
\colhead{(5)} & \colhead{(6)} & \colhead{(7)} & \colhead{(8)}
}
\startdata
G270H   & May 1996  & \cii\ $\lambda$2326    & 3.5$\pm$1.1   & 2.8$\pm$1.3   &  680$\pm$230  & 4400$\pm$1000 & 600$\pm$75 \nl
\tableline
G650L   & ''        & \hb\              & 7.5$\pm$1.8   & 1.13$\pm$0.26\tablenotemark{a} &  963$\pm$75\tablenotemark{a}   & 3060$\pm$260\tablenotemark{a}  & 560$\pm$130 \nl
''      & ''        & \oiii\ $\lambda$4959  & 9.2$\pm$2.0   & ''            &  ''           & ''            & '' \nl
''      & ''        & \oiii\ $\lambda$5007  & 19.3$\pm$2.1  & ''            &  ''           & ''            & '' \nl
\tableline
G780H   & ''        & \oi\ $\lambda$6300    & 11.7$\pm$1.7  & 1.13$\pm$0.26 & 963$\pm$75    & 3060$\pm$260  & 554$\pm$20 \nl
''      & ''        & \oi\ $\lambda$6363    & 4.5$\pm$1.0   & ''            & ''            & ''            & '' \nl
''      & ''        & \nii\ $\lambda$6548   & 21.9$\pm$2.7  &  ''            & ''            & ''            & '' \nl
''      & ''        & \nii\ $\lambda$6584   & 65.7$\pm$8.1\tablenotemark{b} & ''            & ''            & ''            & '' \nl
''      & ''        & \ha\             & 20.4$\pm$3.9  & ''            & ''            & ''            & '' \nl
''      & ''        & \sii\ $\lambda$6716   & 18.1$\pm$2.8  & ''            & ''            & ''            & '' \nl
''      & ''        & \sii\ $\lambda$6731   &  9.6$\pm$2.2  & ''            & ''            & ''            & '' \nl
\tableline
G570H   & Jan. 1997 & \hb\              & 5.08$\pm$0.35 & 2.07$\pm$0.18 & 731$\pm$40    &  4350$\pm$230 & 147$\pm$14 \nl
''      & ''        & \oiii\ $\lambda$4959  & 4.40$\pm$0.39 & ''            & ''            & ''            & '' \nl
''      & ''        & \oiii\ $\lambda$5007  & 15.00$\pm$0.89 & ''           & ''            & ''            & '' \nl
\tableline
''      & ''        & \oi\ $\lambda$6300    & 7.63$\pm$0.67 & 1.90$\pm$0.22 & 714$\pm$35    & 2700$\pm$100  & 281$\pm$8\tablenotemark{d} \nl
''      & ''        & \oi\ $\lambda$6363    & 2.90$\pm$0.42 & ''            & ''            & ''            & '' \nl
''      & ''        & \nii\ $\lambda$6548   &  17.9$\pm$1.3 & ''            & ''            & ''            & '' \nl
''      & ''        & \nii\ $\lambda$6584   &  53.7$\pm$3.9\tablenotemark{b} & ''           & ''            & ''            & '' \nl
''      & ''        & \ha\             &  12.7$\pm$1.6 & ''            & ''            & ''            & 48$\pm$44\tablenotemark{c} \nl
''      & ''        & \sii\ $\lambda$6716   &  11.9$\pm$1.2 & ''            & ''            & ''            & 281$\pm$8\tablenotemark{d} \nl
''      & ''        & \sii\ $\lambda$6731   &  8.6$\pm$1.0  & ''            & ''            & ''            & '' \nl
\enddata
\tablenotetext{a}{Since the S/N of this spectrum is low, these values
    have been held fixed at the values determined in the G780H
    spectrum.}
\tablenotetext{b}{The ratio to the two \nii\ lines has been forced to be
    exactly equal to the theoretical value 3.0.}
\tablenotetext{c}{A proper fit to the \ha+\nii\ complex could only
    be obtained by allowing the \ha\ components to have a different
    redshift from the other emission lines.}
\tablenotetext{d}{These values have been constrained to be the same.}
\tablecomments{Col.~(1): FOS grating used in the observations; Col.~(2):
    Date the observation was taken; Col.~(3): Identification of the
    emission line; Col.~(4): Integrated flux in the Narrow Line Region
    (NLR) emission component in units of 10$^{-16}$ erg s$^{-1}$
    cm$^{-2}$; Col.~(5): Ratio between the flux in the Broad Line Region
    (BLR) and NLR emission components. The '' in this column/line and
    others means that the parameters have been constrained to be the
    same as the last value specifically defined in this column;
    Col.~(6): Measured Gaussian intrinsic velocity dispersion of the NLR
    line in units of km s$^{-1}$; Col.~(7): Measured Gaussian intrinsic
    velocity dispersion of the BLR line in units of km s$^{-1}$;
    Col.~(8): Observed redshift in km s$^{-1}$.}
    \end{deluxetable}

\end{document}